\documentclass[]{interact}

\usepackage[T1]{fontenc}
\usepackage[utf8]{inputenc}
\usepackage[english,polish]{babel}
\usepackage{lmodern}       % DODAJ – żeby były wszystkie polskie znaki

\usepackage{array}
\usepackage{makecell}
\usepackage{epstopdf}
\usepackage[caption=false]{subfig}
\usepackage{float}
\usepackage{graphicx}
\usepackage{booktabs}
\usepackage{placeins}
\usepackage{comment}
\usepackage{tabularx}
\usepackage{caption}
\usepackage{xcolor}
\usepackage{url}
\usepackage{hyperref}
\usepackage[numbers,sort&compress]{natbib}

\bibpunct[, ]{[}{]}{,}{n}{,}{,}

\theoremstyle{plain}

\theoremstyle{definition}

\theoremstyle{remark}

\begin{document}
\selectlanguage{english}
\title{Modeling interest rate swap volatility with GARCH processes}

\author{%
  \name{Micha\l{} Balcerek\textsuperscript{a} and Micha\l{} Wronka\textsuperscript{a}\thanks{CONTACT Micha\l{} Wronka. Email: michal.wronka@pwr.edu.pl}}%
  \affil{\textsuperscript{a} Faculty of Pure and Applied Mathematics, Wroc{\l}aw University of Science and Technology, Wyspia\'nskiego 27, 50-370 Wroc{\l}aw, Poland;}%
}

\maketitle

{

\begin{abstract}
We examine the conditional volatility dynamics of the USD 1Y$\times$10Y forward swap rate using GARCH(1,1), GJR-GARCH(1,1), 
and a two-regime Markov-switching GARCH (MSGARCH) model. The analysis uses daily data from 2007 to 2023 and incorporates market-implied 
measures (ATM swaption volatility and the SRVIX index) together with a broad set of diagnostic tests. Standard GARCH and GJR-GARCH models 
show stable short-run parameters, but the intercept $\omega$ varies markedly across rolling windows, causing instability in the implied long-run variance.
 This pattern, confirmed by the Nyblom test, motivates adopting a regime-switching specification.
  MSGARCH mitigates this issue by keeping regime-specific parameters stable and capturing time variation through filtered regime probabilities. 
  It delivers the highest log-likelihood and lowest AIC, whereas BIC favours the more parsimonious GJR-GARCH. One-step-ahead backtesting
   indicates comparable short-horizon accuracy across models, but MSGARCH offers a clearer structural interpretation by isolating high- and low-volatility 
   regimes aligned with major market events.

\end{abstract}

}


\tableofcontents

\section{Introduction}
\label{sec:intro1}

The Generalized Autoregressive Conditional Heteroscedastic (GARCH) models feature widespread usage in the financial industry, especially in equity and FX with the purpose of filtering and forecasting volatility. The model has been developed by Tim Bollerslev in 1986 \cite{Bollerslev}, as an extension of the Autoregressive Conditional Heteroskedasticity (ARCH) model introduced by Engle in 1982 \cite{Engle}. These models were designed to address problems in financial time series data where the volatility varies over time, exhibiting periods of high and low variance. Since its introduction, many extensions of the original GARCH have been developed to address specific features of time series data:
\begin{itemize}
\item{EGARCH, i.e., Exponential GARCH}: introduced by Nelson in 1991 \cite{Nelson}. The model describes logarithmic variance and captures asymmetries. 
\item{TGARCH, i.e., Threshold GARCH, or GJR-GARCH}: Developed by Zakoian in 1994 \cite{Zakoian}  and Glosten, Jagannathan and Runkle in 1993 \cite{Glosten}. It captures different behaviors in volatility depending on the sign of past errors
\item{GARCH-M, i.e., GARCH in Mean}: Proposed by Engle, Lilien and Robins in 1987 \cite{EngleLilienRobins}. It uses conditional variance as a regressor in the mean equation, capturing the risk-return trade-off.
\item{Multivariate GARCH}: This is an extension of univariate GARCH, allowing for modeling of time-varying covariances.
{\item {Markov-switching GARCH (MSGARCH)} : MSGARCH model combines standard GARCH dynamics with a latent Markov chain governing transitions between volatility regimes. Originally introduced by Hamilton \cite{Hamilton1989}
for regime switching and subsequently adapted to conditional volatility by Haas et al. \cite{HaasMittnikPaolella2004}.}
\end{itemize}

Areas of application with respect to subjects such as:
\begin{enumerate}
    \item {Modeling of financial markets}
    \begin{itemize}
    \item{Volatility Forecasting}: GARCH models are extensively used to forecast future volatility of asset prices, which is of paramount importance for risk management, option pricing, and portfolio optimization. See, for instance result obtained by Burnecki, Gajda and Sikora \cite{Burnecki_Gajda_Sikora}, where authors discuss lack of memory in Hang Seng index logarithmic returns as well conditional heteroscedasticity property in the index data.
    \item{Value at Risk (VaR) Estimation}:  Financial institutions use GARCH models to generate scenarios for risk factors underlying estimation of VaR, a measure of the potential loss in value of a portfolio over a defined period for a given confidence interval. See \cite{Barone_Adesi} for Filtered Historical Simulation approach towards modeling of financial risk risk and subsequent Value at Risk calculations.
    \item{Derivatives Pricing}: GARCH models are used to model the underlying asset volatility in option pricing models.
    \end{itemize}
    \item{Macroeconomics}
    \begin{itemize}
    \item{Inflation and Interest Rates}: GARCH models help in analyzing and forecasting the volatility of macroeconomic variables like inflation rates, interest rates and Gross Domestic Product (GDP) which are crucial for economic policy-making.
    \end{itemize}
    \item{Insurance and Actuarial Science}
    \begin{itemize}
        \item{Claims modeling}: In the insurance industry, GARCH models are applied to model the volatility of insurance claims over time, aiding in premium setting and risk assessment.
    \end{itemize}
    \item{Energy markets}
    \begin{itemize}
    \item{Commodity prices}: GARCH models are used to forecast the volatility of commodity prices, such as oil and natural gas, which is important for energy companies and traders in managing price risks. Interesting results were obtained by Janczura and Puć, where the authors proposed dynamic, risk management  strategies for wholesale energy market participants leveraging ARX-GARCH type of models, see \cite{Janczura_Puc}.
    \end{itemize}
   
\end{enumerate}

Volatility projections play a vital role in risk management and derivatives pricing, hence robust volatility modelling is a key requirement for a financial institution exposed to risks and obliged by the regulators to properly account for the capital management.

{
In this article we analyze the volatility dynamics of the USD 1Y$\times$10Y forward
swap rate using a sequence of increasingly flexible conditional heteroskedasticity
models. The structure of the paper is as follows.

\begin{itemize}
    \item Section~\ref{sec:StylizedFacts} introduces the concept of volatility and 
    summarizes the key stylized facts relevant for fixed-income markets, including 
    volatility clustering, fat tails and the impact of monetary policy regimes.

    \item Section~\ref{sec:DataUsed} describes the data set, which consists of daily fair swap 
    rates and European swaption implied volatilities used in the analysis.

    \item Section~\ref{sec:GarchFamily} presents the formal specification of the 
    GARCH(1,1), GJR-GARCH(1,1) and MSGARCH(1,1) models and discusses their relevance for 
    interest rate markets, including the role of leverage-type asymmetry in 
    capturing the negative relationship between changes in swap rates and implied 
    swaption volatility.

    \item Section~\ref{sec:modeltest} evaluates the adequacy of these models using a 
    comprehensive diagnostic framework, including the Nyblom parameter 
    stability test, Engle--Ng sign and size bias tests for asymmetry, and adjusted 
    Pearson goodness-of-fit tests for alternative innovation distributions.

    \item Section~\ref{sec:msgarch} motivates and develops MSGARCH specification. The model allows the conditional variance to evolve according to 
    multiple latent regimes governed by a Markov transition matrix. Each regime 
    has its own GARCH dynamics, enabling the model to capture shifts in volatility 
    persistence, shock sensitivity and tail behaviour arising from structural breaks in 
    interest rate markets (including the LIBOR$\rightarrow$OIS$\rightarrow$SOFR 
    transition).

    \item Section~\ref{sec:backtest} performs an out-of-sample backtest in which models 
    are re-estimated on rolling windows. One-day-ahead volatility forecasts are used 
    to construct 95\% prediction intervals, and empirical coverage statistics are 
    reported.

    \item Section~\ref{sec:regime_effects} discusses the impact of collateralisation regimes
    on  the GARCH and MSGARCH volatility filters and behavior of their parameters. We provide here a comprehensive analysis
of how the GARCH-type parameters evolve across these structural breaks and how
regime changes influence volatility persistence and shock sensitivity.

    \item Section~\ref{sec:TotalVar} compares model-implied volatility filters with market-based measures of total variance, 
    in particular the SRVIX index. Both standard GARCH-type models and MSGARCH produce volatility paths closely aligned with SRVIX, 
    even during stressed periods such as the COVID-19 shock. 
    The similarity in performance reflects the fact that SRVIX aggregates information across the entire volatility surface, 
    while conditional variance filters capture its dominant dynamics. MSGARCH does not materially outperform single-regime models in short-horizon tracking,
     but it provides a clearer structural decomposition of low- and high-volatility phases.

    \item Section ~\ref{sec:SRVIX_contribution} positions SRVIX within the existing literature as an interest-rate analogue of the equity VIX, 
    highlighting its limited use as a benchmark for parametric volatility models. 
    The study contributes by systematically linking GARCH-type conditional variance filters of ATM forward swap rates to 
    SRVIX over a long sample spanning multiple collateralisation regimes. By validating models against both ATM implied swaption volatility and SRVIX, 
    the analysis benchmarks conditional volatility against a measure of total market variance rather than a single point on the volatility surface. 
    This dual validation framework strengthens the empirical relevance of GARCH and MSGARCH models in fixed-income markets.

    \item Section~\ref{sec:concl} summarises the findings and compares the 
    standalone and relative performance of all models considered.
\end{itemize}

The overarching objective of this study is to assess whether standard GARCH and
GJR-GARCH specifications are adequate for modelling interest rate swap volatility,
and to what extent multi-regime extensions such as MSGARCH are required to 
capture the structural breaks, long-run shifts and heavy-tailed behaviour 
characteristic of modern fixed-income markets.

}

\section{Stylized facts about volatility}
\label{sec:StylizedFacts}
In this section we will deal with most common facts and features about volatility, which means everything that could be observed in the  markets as to price variability is concerned. 

\subsection{Definitions of volatility}
\label{sec:VolDef}

From the first principles volatility is a measure of price variability over some period of time. Practitioners often refer to different kinds of volatility. Some of them might be abstract quantities, some are statistical and others are measures of market's assessment of volatility. \newline
We can differentiate among the following types of volatility \cite{Wilmott_introduces}:

\begin{itemize}
\item \textbf{Actual/Local volatility}
\newline
    It may also be considered as a parameter that captures the amount of randomness in an asset's return at any particular instant of time. 
    This is rather difficult to measure but it is used in the Black-Scholes equation.

\item \textbf{Historical/Realized volatility} \newline
    This is a standard error of the set of previous returns. For $n$ trading periods, and returns $r_{t-n}$\dots,$r_{t-1}$ whose average is $\bar{r}$, (see \cite{taylor1})
 \begin{align}\label{eq:stddev} 
        s_{t} = \sqrt{\frac{1}{n-1}\sum_{i=1}^{n}(r_{t-i}-\bar{r})^{2}}.
\end{align}
\newline
Realized volatility might be sometimes used as an estimate of future volatility.

\item \textbf{Implied volatility}
\newline
This is the volatility which, when inserted into the Black-Scholes option pricing formulae, returns the market value of an option and is often referred to as what the market thinks of the future/actual volatility. It might be also impacted by supply and demand \cite{Wilmott_introduces}.

\item \textbf{Stochastic volatility}
\newline
A volatility might be a stochastic variable with its own stochastic differential equation (SDE). Since volatility is known not to be constant it might be interesting to specify how it changes in time. Stochastic  volatility models such as Heston or Stochastic Alpha Beta, Rho (SABR) are examples of continuous time modeling whereas GARCH provides consideration of conditional variance in discrete time.

\item \textbf{Conditional volatility} 
\newline
 Conditional volatility is the standard deviation of a future return that is conditional on known information such as the history of previous returns \cite{taylor1}. The projection for the next period is computed using a discrete time-series model that has been calibrated using appropriate data. In this article, we deal with GARCH family of models which feature convenient and accurate equations for volatility expectations. The autoregressive, conditional heteroscedastic models specify conditional variance $\sigma_{t}$
  of the return in period $t$ using prior information set $I_{t-1}$, where $I_{t-1}$ contains prior returns i.e. ${r_{t-1}, r_{t-2}, \ldots}$

\end{itemize}

\subsection{Variability of volatility}
\begin{figure}[H]
    \centering
    \includegraphics[width=1.0\linewidth]{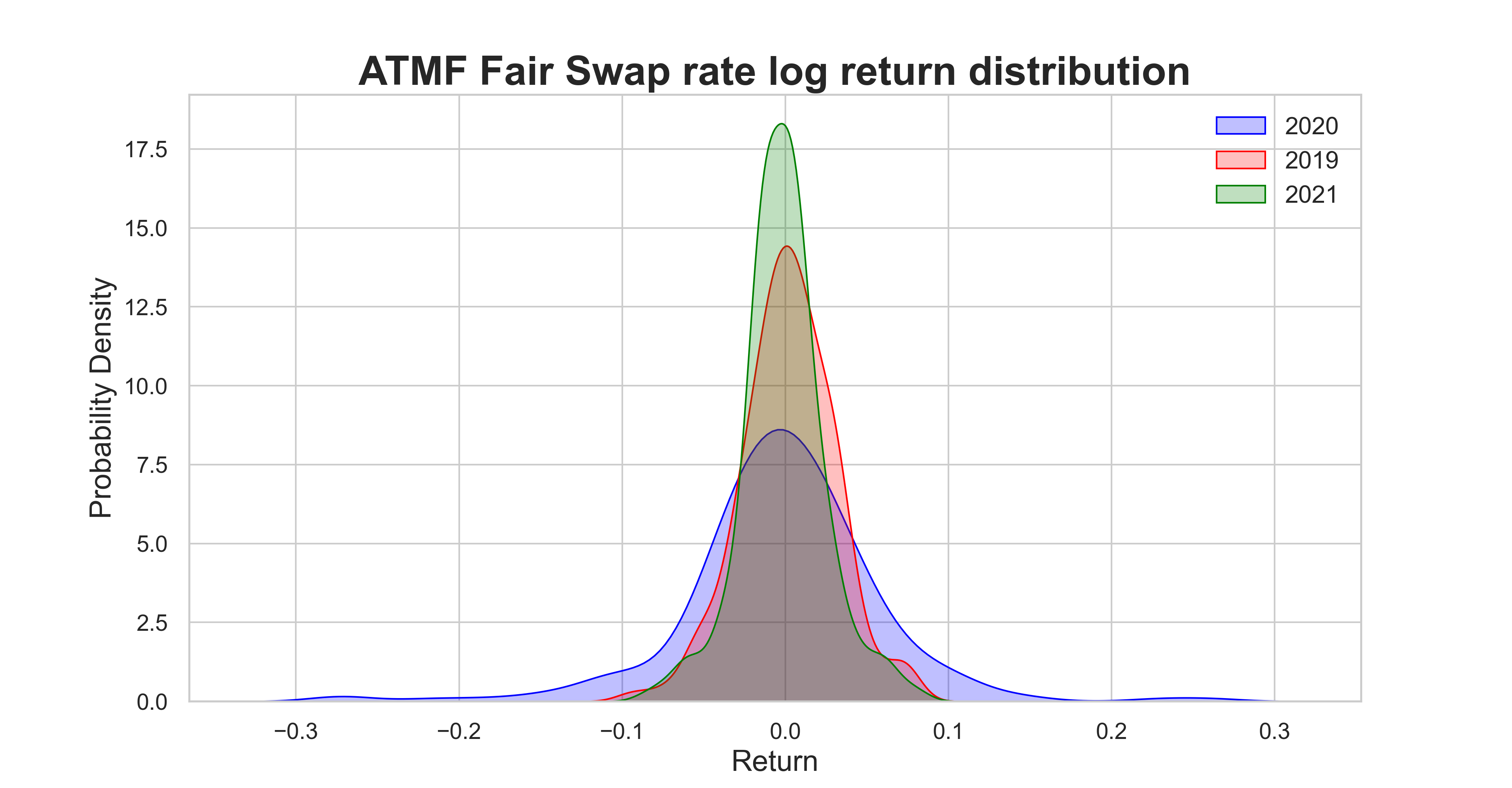}
    \caption{Distribution of Swap Rate log return for 2019, 2020 and 2021.}
    \label{fig:SwapRateChangesDist1}
\end{figure}

Volatility in financial markets varies over time. This observation contradicts assumptions of the Black-Scholes model on constant volatility.  Figure \ref{fig:SwapRateChangesDist1} presents the distribution of changes of the 1Y into 10Y fair swap rate.

\begin{figure}[H]
    \centering
    \includegraphics[width=1.0\linewidth]{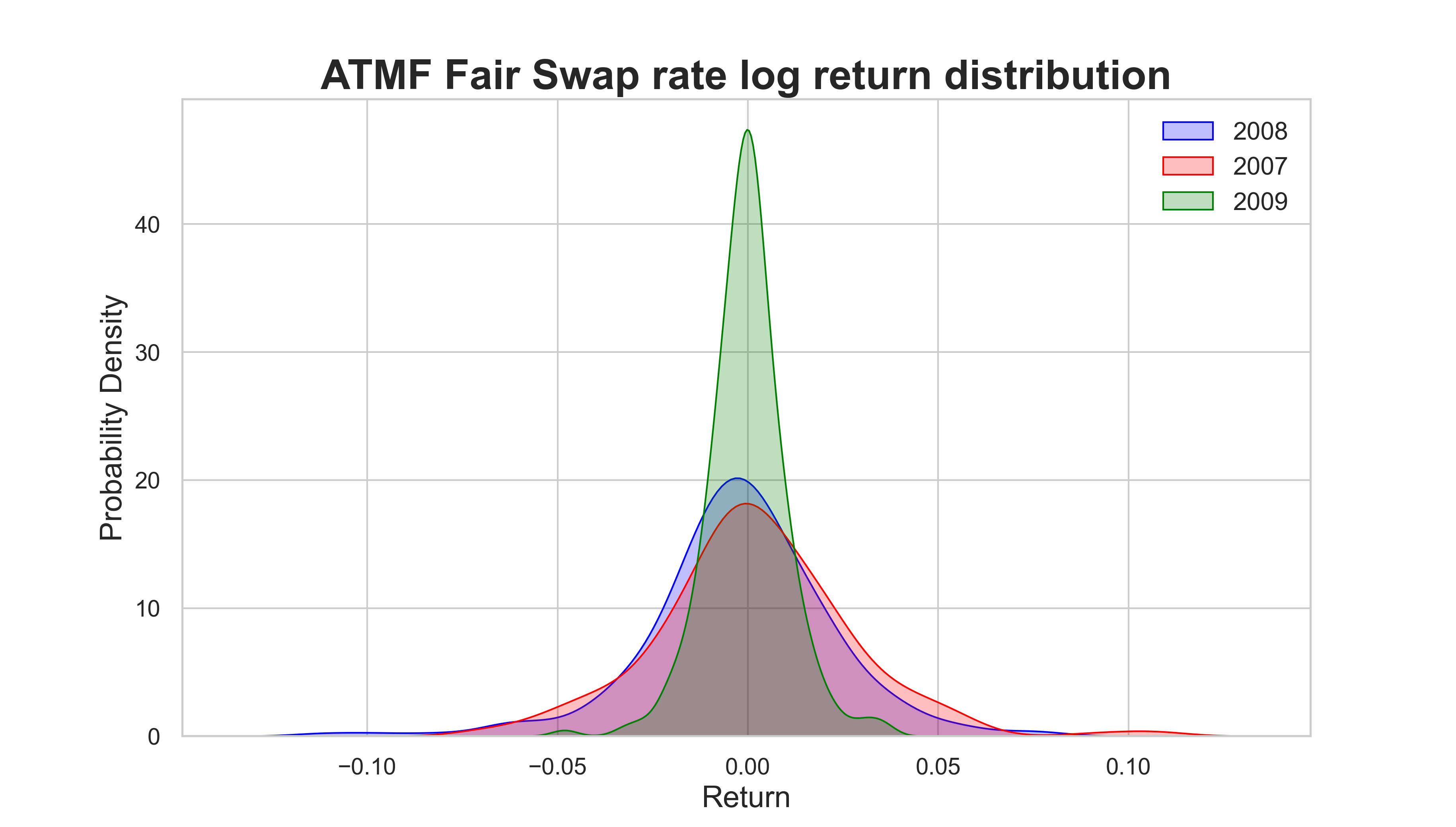}
    \caption{Distribution of Swap Rate log return for 2007, 2008 and 2009.}
    \label{fig:SwapRateChangesDist2}
\end{figure}

In the first year of the COVID-19 pandemic we can see heavier tails indicating higher volatility than against both 2019 and 2021.
A similar situation occurred before in 2008 when financial markets were already very nervous because of the rising subprime crisis and the eventual collapse of Lehman Brothers, see Figure \ref{fig:SwapRateChangesDist2}. In this case, we can see heavier tails already in 2007 and in 2008 as well.

\subsection{Volatility clustering}
\begin{figure}[H]
    \centering
    \includegraphics[width=1.0\linewidth]{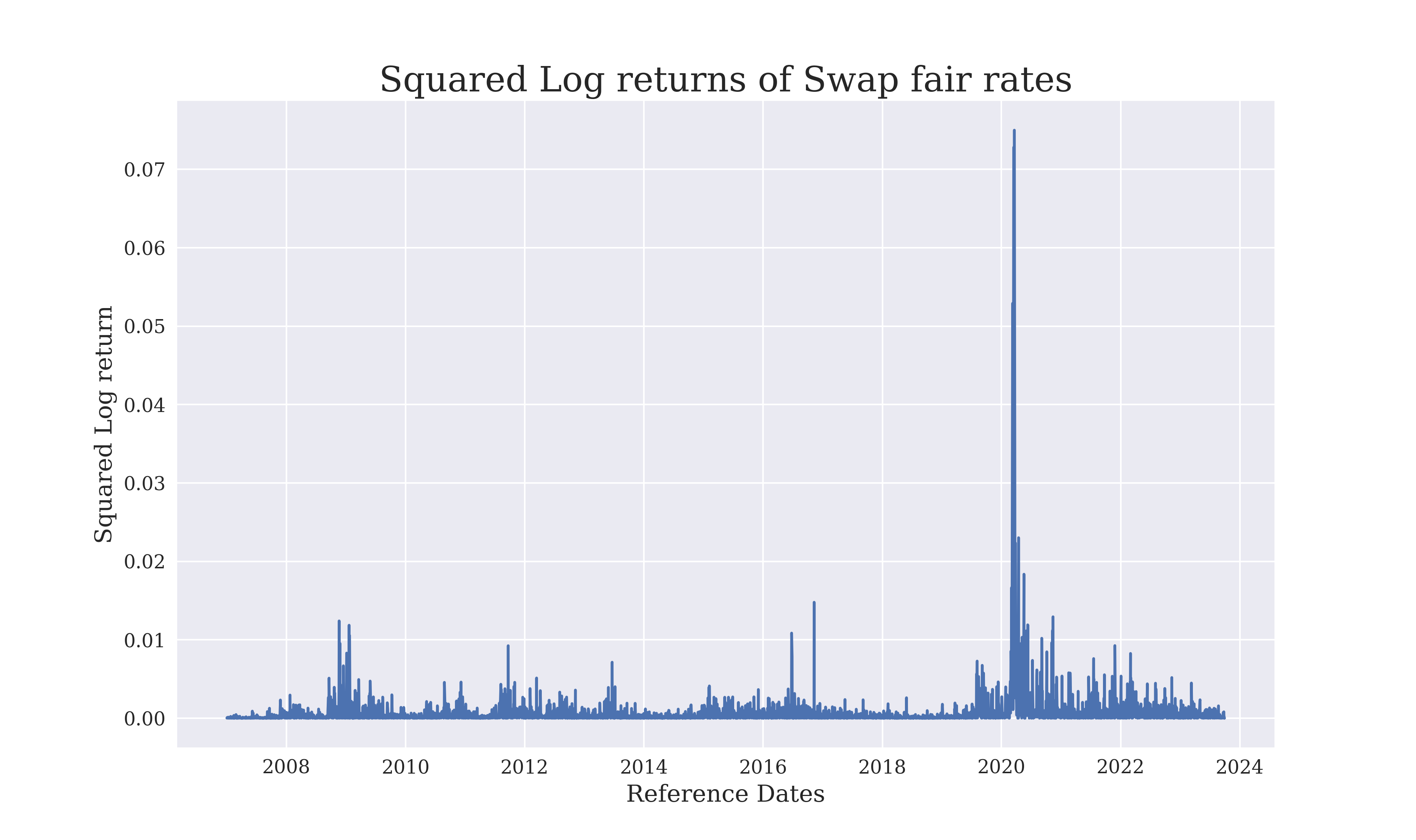}
    \caption{Squared log returns in the time range from 2007 to 2023.}
    \label{fig:SquaredLogReturns}
\end{figure}

Figure \ref{fig:SquaredLogReturns} presents squared log-returns of the swap fair returns. One can clearly see periods with lower and higher values respectively. 
Figure \ref{fig:SquaredLogReturns} shows that volatile periods tend to be followed by other volatile periods. Hence, periods of lower and higher volatility seem to be extended over time and are often accompanied by a sudden shift.
Intuition says that markets need time to digest big news, hence volatility persists. Statistically, time-varying conditional variance implies that big volatility today may lead to big volatility tomorrow.

\subsection{Mean reversion}
The volatility process itself is usually stationary and mean-reverting. We do not expect volatility to explode and after some time it should come back to its long-run levels. The speed of mean reversion is rather slow and may vary among asset classes.

\subsection{Negative correlation with the underlying}

\begin{figure}[H]
    \centering
    \includegraphics[width=1.0\linewidth]{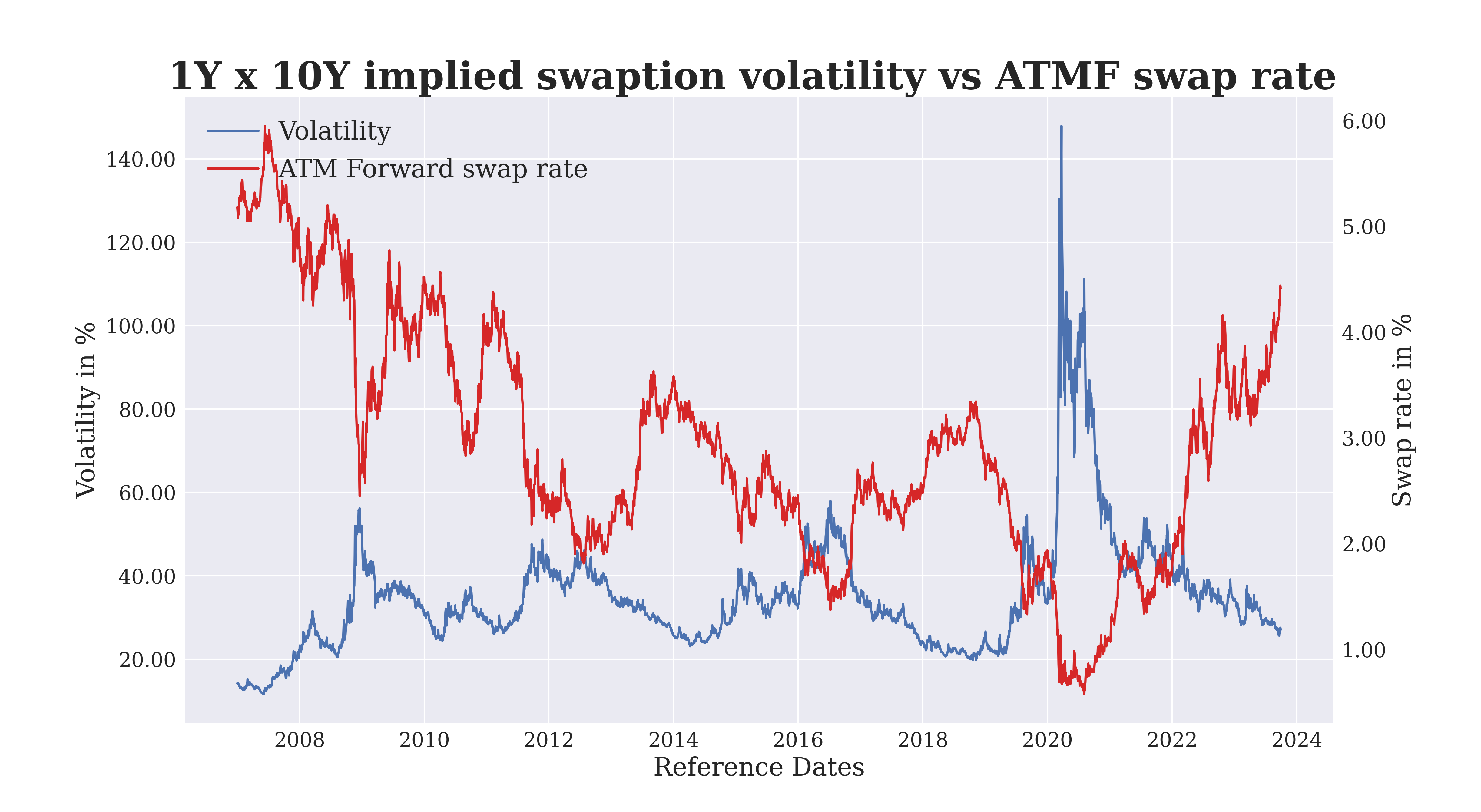}
    \caption{1Y x 10Y implied swaption volatility vs ATMF swap fair rate.}
    \label{fig:SwapRates_vs_Vols_history}
\end{figure}

Volatility is negatively correlated with the returns of the underlying instrument. This is a common feature across asset classes. However, in some cases, e.g., equity indices, it seems more pronounced.
Figure \ref{fig:SwapRates_vs_Vols_history} plots log changes of the underlying swap rates (1Y into 10Y) against ATM swaption volatilities. 

A negative correlation between ATMF swap rates and implied ATM European swaption volatilities is apparent and amounts to $-0.72$ in the examined period. Major drivers of rate movements, such as Lehman’s collapse, the financial crisis, quantitative easing, Brexit, COVID-19, and post-pandemic inflation, are readily apparent.%One can easily identify major drivers of the rates move such as Lehman's collapse, financial crisis, quantitative easing, Brexit, COVID-19 and post-pandemic inflation.

\section{Data used in analysis}
\label{sec:DataUsed}

For this analysis, all relevant data has been purchased from external data provider, i.e., S\&P Global, a London-based vendor company that provides market data and analytics to the broad spectrum of market participants and investors, see \cite{spglobal}.
The data package starts on January 1st 2007 and ends on September 29th 2023. The content of the data applies to the following items:
\begin{itemize}
    \item{Historical data referring to the USD at-the-money forward swap fair rates across three collateralization regimes i.e. LIBOR, OIS and SOFR.}
    \item {Historical data referring to the lognormal implied European Swaption volatilities, both with respect to ATM as well as predefined smile spread strikes.}
    \item {Historical data referring to the normal implied European Swaption volatilities, both with respect to ATM as well as predefined smile spread strikes.}
    \item {Historical data referring to the calibrated SABR parameters to the provided volatility smile.}
    \item {Historical data referring to the calibrated zero rates of the USD SOFR curve.}
\end{itemize}
The purchased data is not raw market data, it has been calibrated and processed based on internal expertise and proprietary models. Such data is not publicly available and requires payment for access. This highlights both the proprietary nature and exclusivity of the data, as well as the cost associated with obtaining it.

\begin{figure}[H]
    \centering
    \includegraphics[width=1.0\linewidth]{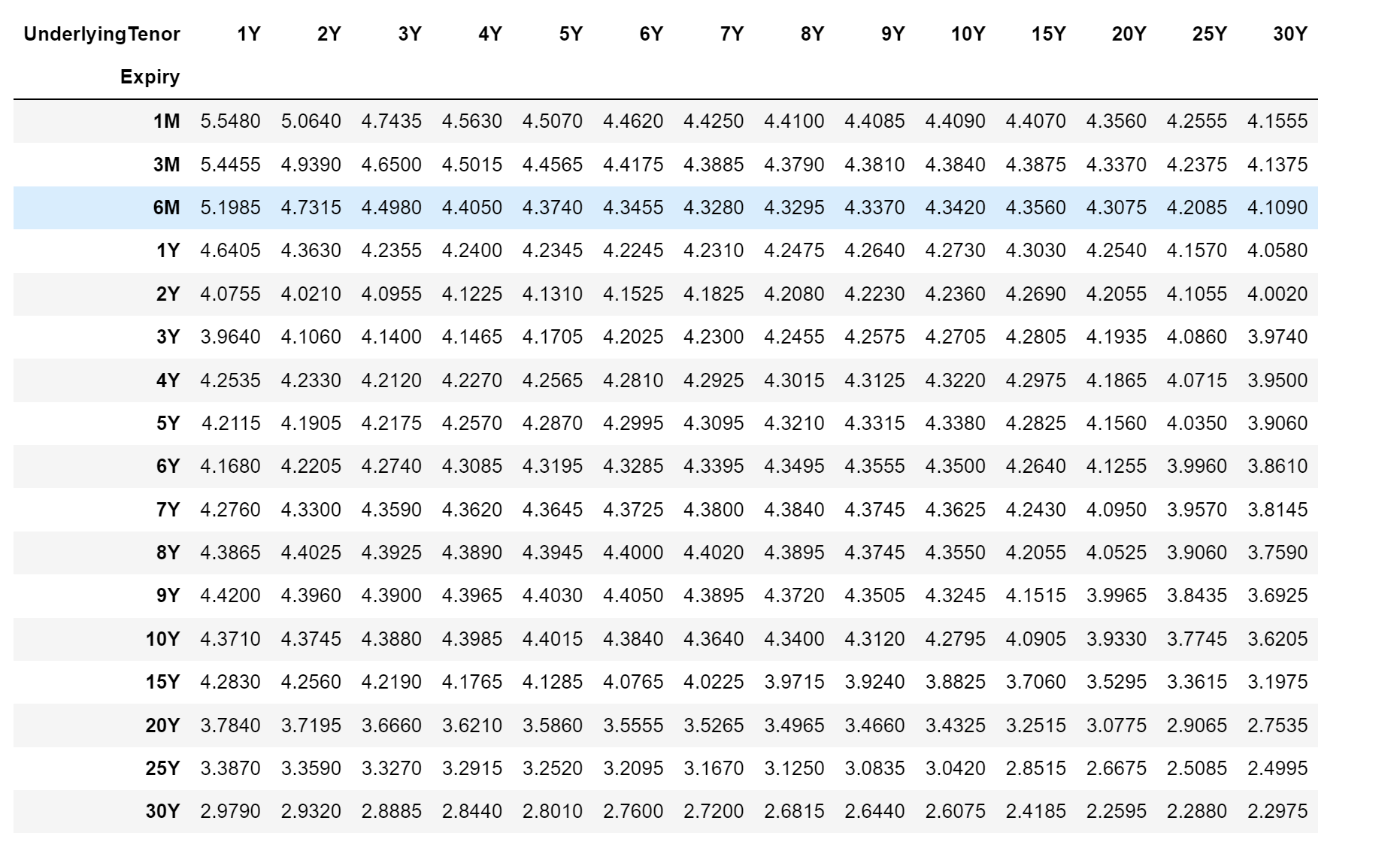}
    \caption{At-the-money forward swap fair rates as of September 29th 2023. Header row contains swap tenors and the first column swaption expiry.}
    \label{fig:FairSwapRates}
\end{figure}

The analysis leverages on at-the-money forward swap fair rates and treats these as market observable values and lognormal implied European Swaption volatilities. We had a dilemma as to which expiry and tenor combination regarding swap data for the historical data to use.  The natural choice seems to be the one that is most liquid. 
Figure \ref{fig:FairSwapRates} contains 238 entries each representing fair rate of a forward starting swap with row index denoting moment of swap start (measured in months or years from 29th of September 2023) and column denoting length of the swap in years. 

Analyzing data available in the Internet regarding the trading volume of European Swaptions \cite{swaptionvolumes} in 2021 and information gathered from the Chicago Board of Options Exchange \cite{SrVix}, we have decided to select a swap that starts in 1Y and ends in 10Y, i.e., 1x10.

\newpage

\section{GARCH family models formulation}
\label{sec:GarchFamily}

This section introduces the volatility models examined throughout the paper, beginning with the standard GARCH framework and its asymmetric extension,
the GJR-GARCH specification. We outline the key structural features of each model, explain how conditional variance responds to new information,
and discuss the role of the innovation distribution in capturing heavy tails and asymmetry in interest rate movements. The section also motivates the transition
to regime-dependent volatility modelling by highlighting the limitations of single-regime GARCH dynamics when structural breaks or shifts in market conditions
are present. Together, these models form the methodological foundation for the empirical analysis conducted in subsequent Sections.

\subsection{GARCH}
\label{sec:Garch}

GARCH$(p,q)$ is defined as:
\begin{align}
\label{eq:Return}
X_{t} &= \mu + \epsilon_{t}, \\
\epsilon_t &= \sigma_t z_t, \qquad z_t \sim F(0,1), \label{eq:innovations}\\[4pt]
\label{eq:GenGARCH}
\sigma_{t}^{2} &= \omega + \sum_{i=1}^{p}\beta_{i}\sigma_{t-i}^{2} + \sum_{j=1}^{q}\alpha_{j}\epsilon_{t-j}^{2}.
\end{align}

The conditional variance is dependent on $q$ most recent squared residuals and $p$ most recent conditional variances. Many studies showed $p=q=1$ is appropriate, cf. \cite{taylor1}.

Current variance depends on:
\begin{itemize}
    \item {past variances: $\sigma_{t-i}^{2}, i=1,2,...$},
    \item {square of past errors: $\epsilon_{t-j}^{2}, j = 1,2,3,...$}.
\end{itemize}
%\newline
In particular, GARCH(1,1) reads:
\begin{align}\label{eq:GARCH11}
\sigma_t^2 = \omega + \alpha \epsilon_{t-1}^{2} + \beta \sigma_{t-1}^{2},
\end{align}
which is a natural extension of the ARCH(1) specification:
\begin{align}
X_t &= \mu + \epsilon_t, \label{eq:return_eq}\\[4pt]
\epsilon_t &= \sigma_t z_t, \qquad z_t \sim F(0,1), \label{eq:innovations}\\[4pt]
\sigma_t^2 &= \omega + \alpha \epsilon_{t-1}^{2}. \label{eq:ARCH1}
\end{align}

Here, $z_t$ denotes the \emph{innovation}, i.e.\ the standardized shock driving the
conditional variance process. The distribution $F$ can be chosen as standard
normal, Student-\emph{t}, GED or Johnson~SU, depending on the desired tail
behavior and skewness properties. The innovation sequence $\{z_t\}$ is assumed
to be i.i.d.\ with zero mean and unit variance and represents the new
information arriving to the market at time $t$. Parameter $\mu$ denotes mean of ATMF swap rate process.

Parameter $\beta$ measures the persistence of conditional volatility, irrespective of what is going on currently in the markets. Given large $\beta$ we can expect that volatility takes a long time to die out after a turbulent time. The variance must be positive, therefore we need to impose restrictions on parameters, i.e., $\omega, \beta, \alpha >0$. We also want to ensure that the variance process is stationary which implies $\beta + \alpha < 1$. We can think that  $\epsilon_{t}$'s are the news that arrive to the marketplace and impact the prices of securities. Hence, a GARCH filter can be seen as an updating mechanism that is dependent on the size of these news. GARCH model also implies an unconditional (long-run) variance as
\begin{align}\label{eq:GARCH(1,1)_longVar}
\bar{\sigma}^{2} = \frac{\omega}{1-\beta-\alpha}.
\end{align}
After small transformations, we can finally write the variance process as:
\begin{align}\label{eq:GARCH(1,1)_final}
\sigma_{t}^{2} = (1-\beta - \alpha)\bar{\sigma}^{2} +\beta\sigma_{t-1}^{2} + \alpha\epsilon_{t-1}^{2}.
\end{align}
Therefore, there are three major forces that drive the variance process
\begin{itemize}
\item{Past variance $\sigma_{t-1}^{2}$ with weight $\beta$},
\item{News' size $\epsilon_{t-1}^{2}$},
\item{Reversion to the long run variance $\bar{\sigma}^{2}$}.
\end{itemize}
\subsection{GJR-GARCH}
\label{sec:GJR}
GARCH model implies that positive and negative residuals have the same impact on the next variance. This might be improper, especially for equity markets. Asymmetric variance models introduce a variable that separates positive and negative residuals. Hence, GARCH(1,1) can be modified so that it incorporates asymmetry. 
GJR-GARCH(1,1) model uses the following modification
\begin{align}
\sigma_{t}^{2} = \omega + \beta \sigma_{t-1}^{2} + \alpha \epsilon_{t-1}^{2} + \gamma \mathbf{1}_{\epsilon_{t-1<0}} \epsilon_{t-1}^{2}.
\end{align}
Impact of residuals $\epsilon_{t-1}^{2}$ on $\sigma_{t}^{2}$ comes in two ways:
\begin{itemize}
\item {When $\epsilon_{t-1} > 0 $ then squared residuals are multiplied by $\alpha$},
\item{When $\epsilon_{t-1} < 0 $ then squared residuals are multiplied by $\alpha + \gamma$}.
\end{itemize}
Similarly, as in the case of regular GARCH, we have to impose restrictions on parameters to ensure that variance is positive which implies that $\omega, \alpha, \beta >0 $ and the variance process is stationary which results in $\alpha + \beta + \frac{\gamma}{2} < 1$.
Leverage effect that is implied by $\gamma$ can be summarized as follows:
\begin{itemize}
\item{$\gamma = 0$: symmetric effect},
\item{$\gamma > 0$: negative residuals are more destabilizing that positive ones},
\item{$\gamma < 0$: positive residuals are more destabilizing that negative ones}.
\end{itemize}
%\subsection{Calibration of standard GARCH and GJR-GARCH model}
%\label{sec:calib}
For the purpose of parameters estimation we are interested in a maximum likelihood approach. Given a series $\{X_{t}\}_{t=1}^{T}$ that represent returns of the underlying instrument we need a conditional density defined as:
\begin{align}\label{eq:ML}
M(X_{t}|\Theta) = f(X_{t}|G_{t-1}, \Theta).
\end{align}
$\Theta$ is a vector of model parameters and $G_{t-1}$ denotes all the information available at time $t-1$, i.e. : $G_{t-1} = \{X_{t-1}, X_{t-2},\dots\}$.    
For the ease of numerical calibration purposes, we will be finally maximizing the sum of log-likelihood expressed as conditional densities that are assumed to be normal
\begin{align}
M = \sum_{t=1}^{T} \log{f(X_{t}|\Theta)},
\end{align}
where
\begin{align}
f(x_{t}|\Theta) = \frac{1}{\sqrt{2\pi\sigma^{2}}}e^{-\frac{(x_{t}-\mu)^{2}}{2\sigma^{2}}}.
\end{align}
In other words, for every single observation, we are looking for GARCH model parameters that maximize the likelihood that the observation falls into the normal distribution given by the mean and conditional variance. 

{
\subsection{MSGARCH}
\label{subsec:msgarch_specification}

Standard single-regime GARCH models assume that the conditional variance 
follows a stable dynamic throughout the sample. However, the diagnostic 
evidence presented in Section~\ref{sec:modeltest}, and in particular the 
Nyblom parameter stability tests, indicates the presence of structural change 
in long-run variance levels and tail behaviour. This motivates the use of MSGARCH model \cite{Gray1996}, \cite{ArdiaHoogerheide2010},  which allow the volatility process 
to alternate between distinct regimes according to an unobserved Markov chain.

Let $\{S_t\}_{t\ge 1}$ denote a latent first-order Markov chain taking values 
in $\{1,\ldots,K\}$ with transition probabilities
\begin{equation}
P(S_t = j \mid S_{t-1} = i) = p_{ij}, \qquad i,j \in \{1,\ldots,K\},
\label{eq:transition_probs}
\end{equation}
where $P = [p_{ij}]$ is a $K \times K$ transition matrix with $p_{ij} \ge 0$ 
and $\sum_{j=1}^{K} p_{ij} = 1$ for each $i$.

Conditional on regime $S_t = k$, the return dynamics follow
\begin{align}
X_t &= \mu_k + \epsilon_{t}, \\
\epsilon_t &= \sigma_{t,k} z_{t,k}, \qquad z_{t,k} \sim F_k(0,1),
\end{align}
where $F_k$ denotes the innovation distribution associated with regime $k$. In 
our empirical application we consider both Student-\emph{t} and Johnson~$SU$ 
innovations to accommodate heavy tails and skewness observed in standardized 
residuals.

The conditional variance evolves according to a regime-dependent GARCH(1,1) 
process:
\begin{equation}
\sigma_{t,k}^2 = \omega_k + \alpha_k \epsilon_{t-1}^2 + \beta_k \sigma_{t-1,k}^2,
\label{eq:msgarch_variance}
\end{equation}
so that each volatility regime is characterised by its own persistence, 
short-run sensitivity and unconditional variance level. The latent state 
variable $S_t$ determines which parameter vector 
$\theta_k = (\mu_k, \omega_k, \alpha_k, \beta_k, \text{shape}_k)$ is active at 
time $t$. The parameter $\text{shape}_k$ captures the shape of the innovation distribution in
regime $k$ and is not part of the GARCH variance dynamics itself. Instead, it governs
the distributional properties of the standardised innovations $z_t$, in particular
tail thickness and, where applicable, skewness. Its exact interpretation depends on
the assumed distributional family: for the Student-$t$ distribution it corresponds to
the degrees of freedom, for the Generalised Error Distribution it controls departures
from Gaussian kurtosis, and for the Johnson~$SU$ specification it represents a vector
of shape parameters determining both asymmetry and tail behaviour. Allowing the
shape parameter to vary across regimes enables the model to capture differences in
the distribution of shocks between tranquil and stressed market conditions.

The likelihood contribution at each step is obtained by averaging over all 
possible states, weighted by their filtered probabilities:
\begin{equation}
\ell_t = \sum_{k=1}^{K} 
    f_k(X_t \mid \theta_k, \mathcal{F}_{t-1}) \, 
    \P(S_t = k \mid \mathcal{F}_{t-1}),
\end{equation}
and the full-sample log-likelihood is given by 
$\log L = \sum_{t=1}^{T} \log \ell_t$. Parameter estimation proceeds by 
maximising the log-likelihood jointly with respect to the regime-specific 
parameters and the transition probabilities $p_{ij}$, subject to standard 
positivity and stationarity constraints. 

The MSGARCH specification generalises the standard GARCH model in two important 
dimensions: (i) it allows volatility persistence and long-run variance levels 
to differ across regimes, and (ii) it allows the innovation distribution to 
change across regimes. This flexibility is essential in the presence of market 
regime switches such as those induced by the financial crisis of 2008, 
quantitative easing policies, or the recent transition from LIBOR to SOFR. 
Later sections demonstrate that MSGARCH models substantially improve parameter 
stability, better capture tail behaviour, and align more closely with 
market-implied volatility dynamics.
}
{
\subsection{Alternative innovation distributions: GED and Johnson\,\texorpdfstring{$SU$}{SU}}
\label{subsec:dist_GED_JSU}

Standard GARCH models are typically estimated under the assumption of Gaussian
innovations. However, financial return series frequently exhibit excess kurtosis and
asymmetry, which cannot be captured adequately by the normal distribution. For this
reason, we also consider two flexible alternatives: the Generalised Error Distribution
(GED) \cite{Nelson1991} and the Johnson~$SU$ \cite{JohnsonSU} distribution.

\textbf{Generalised Error Distribution (GED)}
The Generalised Error Distribution (GED), introduced by Nelson in the context of 
EGARCH models \cite{Nelson1991}, is a flexible symmetric family whose tail 
thickness is governed by its shape parameter. Depending on the value of this 
parameter, the GED can generate tails that are heavier or lighter than the Gaussian 
benchmark, but it does not constitute a universal heavy-tailed distribution in the 
strict probabilistic sense. In particular, the notion of “heavy tails’’ is not defined 
here in terms of regular variation or power-law decay, and therefore the GED should 
be viewed primarily as a convenient parametric extension for modelling deviations 
from normality rather than a canonical heavy-tailed distribution.

The probability density function of the GED on the real line, $z \in \mathbb{R}$, is given by
\[
f_{\mathrm{GED}}(z;\nu)
=
\frac{\nu}{2^{1+1/\nu}\,\Gamma(1/\nu)}
\exp\!\left(
-
\frac{1}{2}\lvert z\rvert^{\nu}
\right),
\qquad 
\nu>0.
\]
The shape parameter $\nu$ controls tail thickness: $\nu=2$ recovers the Gaussian
distribution, while $\nu<2$ allows for heavier tails. As our diagnostic results in
Section \ref{subsec:gof_summary} indicate, the GED improves the fit relative to the Gaussian specification,
though it remains symmetric and therefore unable to account for skewness in the data.

\textbf{Johnson\,\texorpdfstring{$SU$}{SU} distribution.}
The Johnson~$SU$ distribution is a four-parameter family obtained by applying a 
non-linear transformation to a standard normal random variable. Let 
$Y \sim \mathcal{N}(0,1)$ and define
\[
Z \;=\; \xi + \lambda \, \sinh\!\left(\frac{Y - \gamma}{\delta}\right),
\]
where $\xi \in \mathbb{R}$ is a location parameter, $\lambda > 0$ is a scale parameter,
and $\gamma \in \mathbb{R}$, $\delta > 0$ are shape parameters controlling skewness
and kurtosis. In this construction $Y$ is a standard normal variable taking values on 
the whole real line, $Y \in \mathbb{R}$, while $Z$ is the transformed Johnson~$SU$ 
random variable, also supported on the entire real line, $Z \in \mathbb{R}$.

The corresponding probability density function of $Z$ on $\mathbb{R}$ is
\[
f_{SU}(z; \gamma,\delta,\xi,\lambda)
=
\frac{1}{\lambda \sqrt{2\pi}} \,
\frac{\delta}{\sqrt{1 + \left( \frac{z - \xi}{\lambda} \right)^{2}}}
\,
\exp\!\left\{
-\frac{1}{2}
\left[
\gamma + \delta \, \operatorname{arcsinh}\!\left(\frac{z - \xi}{\lambda}\right)
\right]^{2}
\right\},
\qquad z \in \mathbb{R},
\]
Due to its ability to accommodate both skewness and fat tails, the Johnson~$SU$ distribution achieves the
best overall goodness-of-fit in our analysis presented in Section \ref{sec:modeltest}, even though goodness-of-fit
tests still reject exact distributional correctness at conventional levels.

\medskip
In subsequent sections, we evaluate the performance of all innovation distributions
using likelihood-based criteria (AIC, BIC), Pearson goodness-of-fit diagnostics, and
Engle--Ng asymmetry tests. These results demonstrate that alternative distributions
provide a substantially more accurate representation of interest-rate swap return
dynamics than the Gaussian assumption.
}

\begin{comment}
\subsection{FIGARCH}
\label{sec:FIGARCH}

The FIGARCH (Fractionally Integrated GARCH) model is a type of time series model used to describe the conditional variance of a financial asset's returns. It extends the traditional GARCH model by incorporating a fractional differencing parameter to capture long memory or long-range dependence in the volatility process.

The general formula for the conditional variance 
\begin{align}
\sigma_t^2 = \omega + \sum_{i=1}^{p} \alpha_i \epsilon_{t-i}^2 + \sum_{j=1}^{q} \beta_j \sigma_{t-j}^2 + \sum_{k=1}^{o} \gamma_k (\epsilon_{t-k}^2 - \lambda_k \sigma_{t-k}^2)
\end{align}
  
Where:
\begin{itemize}
    \item \( \omega \) is the constant term.
    \item \( \alpha_i \) and \( \beta_j \) are the coefficients of the ARCH and GARCH terms respectively.
    \item \( \gamma_k \) is the coefficient of the fractional integrated GARCH term.
    \item \( \epsilon_t \) is the error term, typically assumed to be independent and identically distributed (i.i.d.) with mean zero and constant variance.
    \item \( p \), \( q \), and \( o \) are the orders of the ARCH, GARCH, and fractional integrated GARCH terms respectively.
    \item \( \lambda_k \) is the fractional differencing parameter, which determines the degree of persistence or long memory in the volatility process.
\end{itemize}

The fractional integrated GARCH term allows the conditional variance to be influenced by both past squared innovations and past conditional variances, with the influence decaying over time with a fractional differencing 
\end{comment}

\section{Model testing and validation}
\label{sec:modeltest}

In this section we assess whether the proposed GARCH-type specifications are
statistically adequate to describe the dynamics of daily log changes in the
ATMF 1Y$\times$10Y swap rate. We first establish the presence of ARCH effects
and present baseline GARCH and GJR-GARCH calibrations. We then perform a
battery of hypothesis tests on residuals and standardised residuals, followed
by advanced diagnostics covering parameter stability, asymmetry and
goodness-of-fit of innovation distributions.

\subsection{Evidence of ARCH effects and baseline calibration}

First, we test for the presence of conditional heteroscedasticity in the
log changes of ATMF swap rates. A time series exhibiting autocorrelation in
squared observations is said to display an autoregressive conditional
heteroscedastic (ARCH) effect. Engle’s ARCH test is a Lagrange multiplier test
for the null hypothesis of no ARCH effects; see \cite{Matlab_ARCH_Test}.

\begin{figure}[H]
    \centering
    \includegraphics[width=1.0\linewidth]{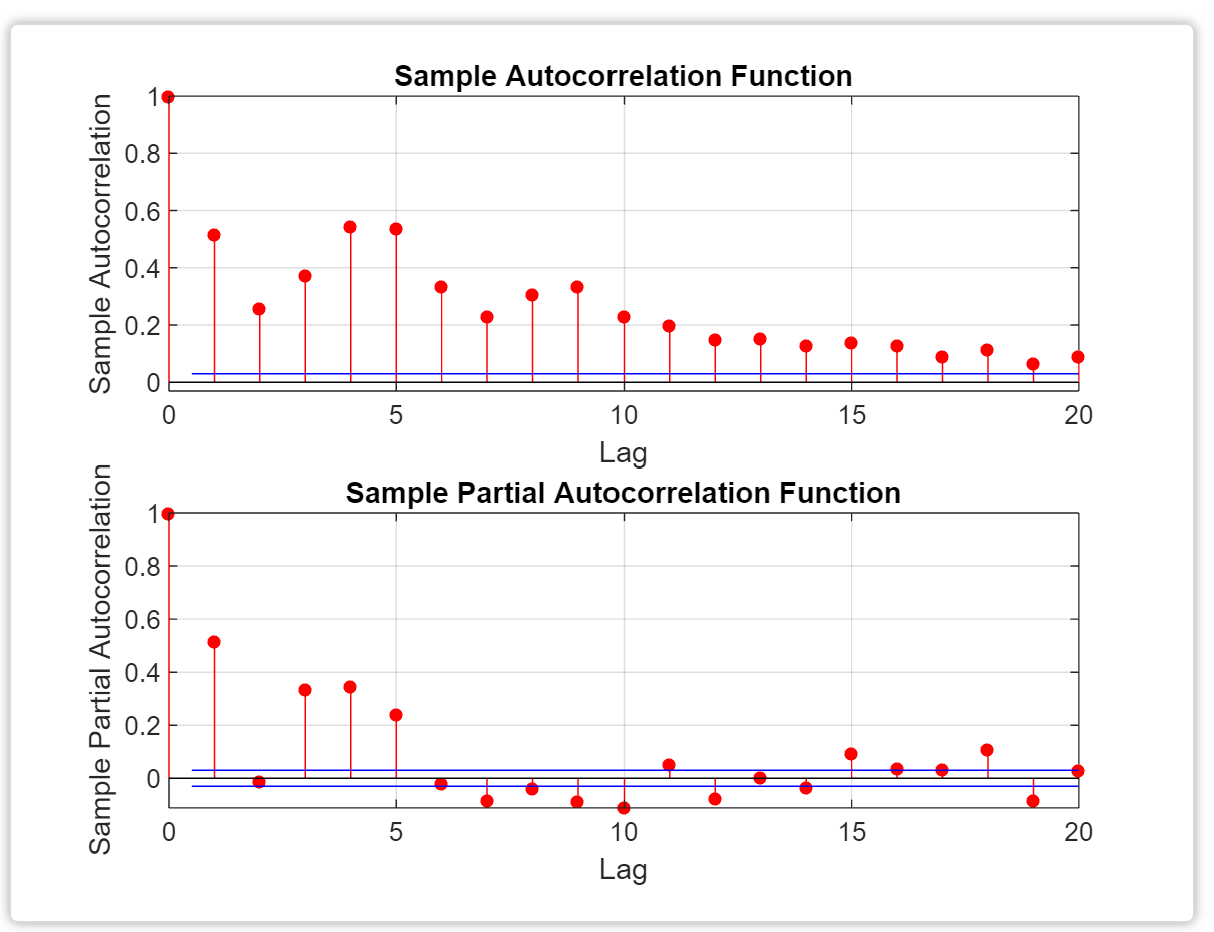}
    \caption{Autocorrelation Function and Partial Autocorrelation Function of squared residuals for ATMF swap rates.}
    \label{fig:ACF_PACF}
\end{figure}

Figure~\ref{fig:ACF_PACF} presents the autocorrelation function (ACF) and
partial autocorrelation function (PACF) of squared residuals.

We observe strong autocorrelation in the squared residuals. At the 5\%
significance level, Engle's test (with default one lag) rejects the null hypothesis of no ARCH effect
in favour of the alternative. Hence, we are fully justified in applying GARCH
models to filter the conditional variance.

\begin{figure}[!ht]
    \centering
    \includegraphics[width=1.0\linewidth]{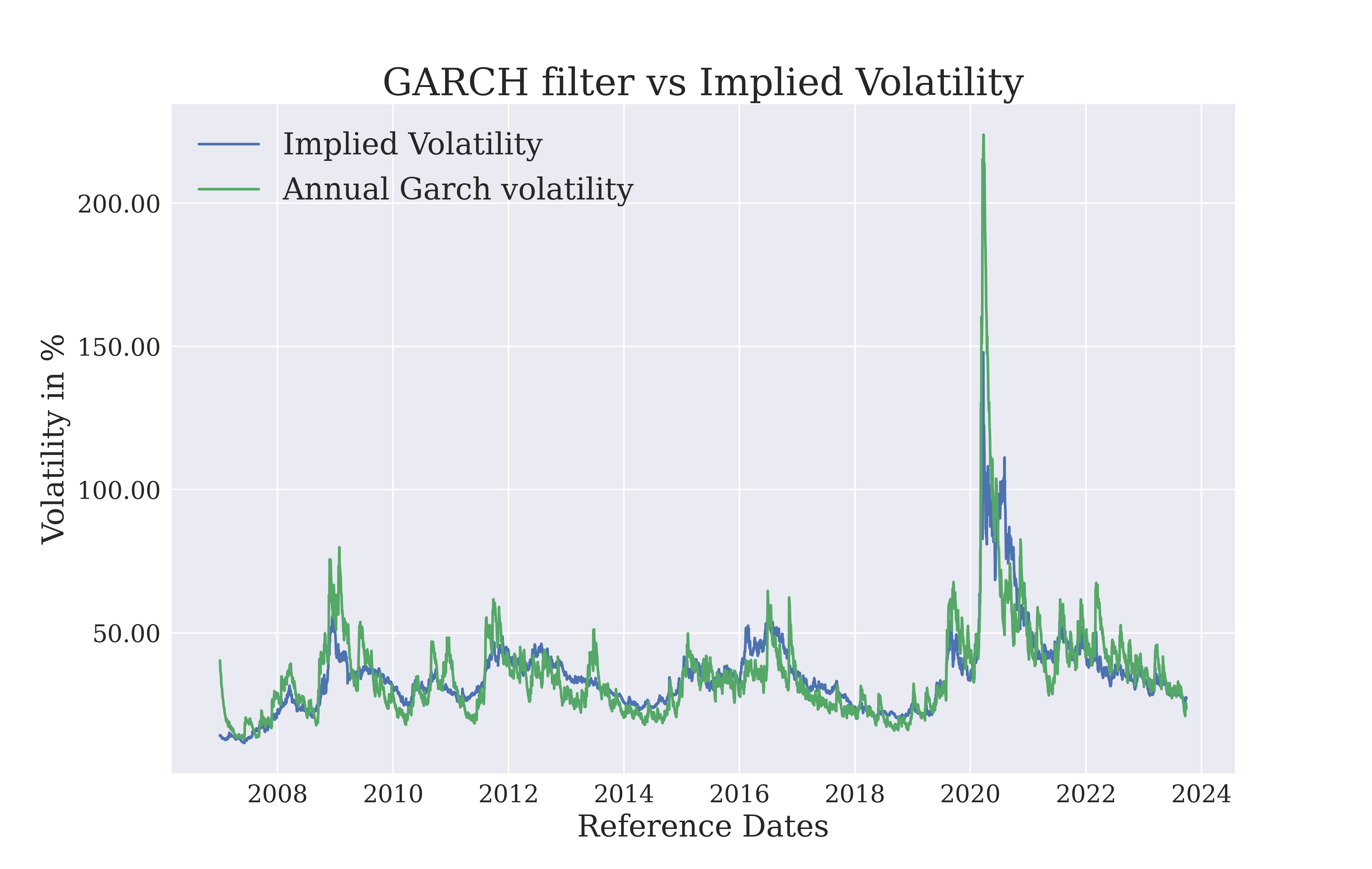}
    \caption{Regular GARCH model calibration results. Blue line denotes market implied ATM swaption volatilities, green one annualized GARCH volatility.}
    \label{fig:REG_GARCH}
\end{figure}

Calibration of the GARCH(1,1) specification confirms the statistical
significance of the parameters $\omega$, $\alpha$ and $\beta$. The resulting
estimates are reported in Table~\ref{tab:GARCH_CALIB}. Figure~\ref{fig:REG_GARCH} presents the conditional annualised volatility
filtered by the GARCH model together with the history of ATMF implied swaption
volatility.

\begin{table}[!ht]
     \caption{GARCH model calibration results, where header line description stands for parameter name, parameter value, standard error, t-Student statistic and p-value respectively.}
     \label{tab:GARCH_CALIB}
    \centering
    \begin{tabular}{|c|c|c|c|c|}\hline
       Parameter & Value & Standard Error  & TStatistic & PValue \\ \hline
         $\omega$ & 0.000003 & 0.0000 & 3.4864 & 0.00049\\ \hline
         $\beta$ & 0.9254 & 0.0053 & 173.69 & $<0.0001$\\ \hline
         $\alpha$ & 0.0723 & 0.0055  & 13.279 & $<0.0001$\\ \hline
    \end{tabular}
\end{table}

\FloatBarrier

The GARCH-filtered volatility tracks ATM implied swaption volatility closely
over the whole sample. At lower volatility levels the GARCH filter tends to
underestimate, whereas during high-volatility episodes it slightly overshoots
market-implied volatility.

Table~\ref{tab:GJR-GARCH_CALIB} reports the calibration results for the
GJR-GARCH(1,1) specification, which allows for asymmetric responses of
volatility to negative and positive shocks. Again, all parameters are highly
significant.

\begin{table}[H]
   \caption{GJR-GARCH model calibration results, where header line description stands for parameter name, parameter value, standard error, t-Student statistic and p-value respectively.}
    \label{tab:GJR-GARCH_CALIB}
    \centering
    \begin{tabular}{|c|c|c|c|c|} \hline
       Parameter & Value & Standard Error  & TStatistic & PValue \\ \hline
         $\omega$ & 0.0000 & 0.0000 & 3.0396 & 0.0024\\ \hline
         $\beta$ & 0.9347 & 0.0052 & 178.95 & $<0.0001$\\ \hline
         $\alpha$ & 0.0436 & 0.0059 & 7.3864 & $<0.0001$ \\ \hline
         $\gamma$ & 0.0411 & 0.0071  & 5.83 & $<0.0001$ \\ \hline
    \end{tabular}
\end{table}

\begin{figure}[H]
    \centering
    \includegraphics[width=1.0\linewidth]{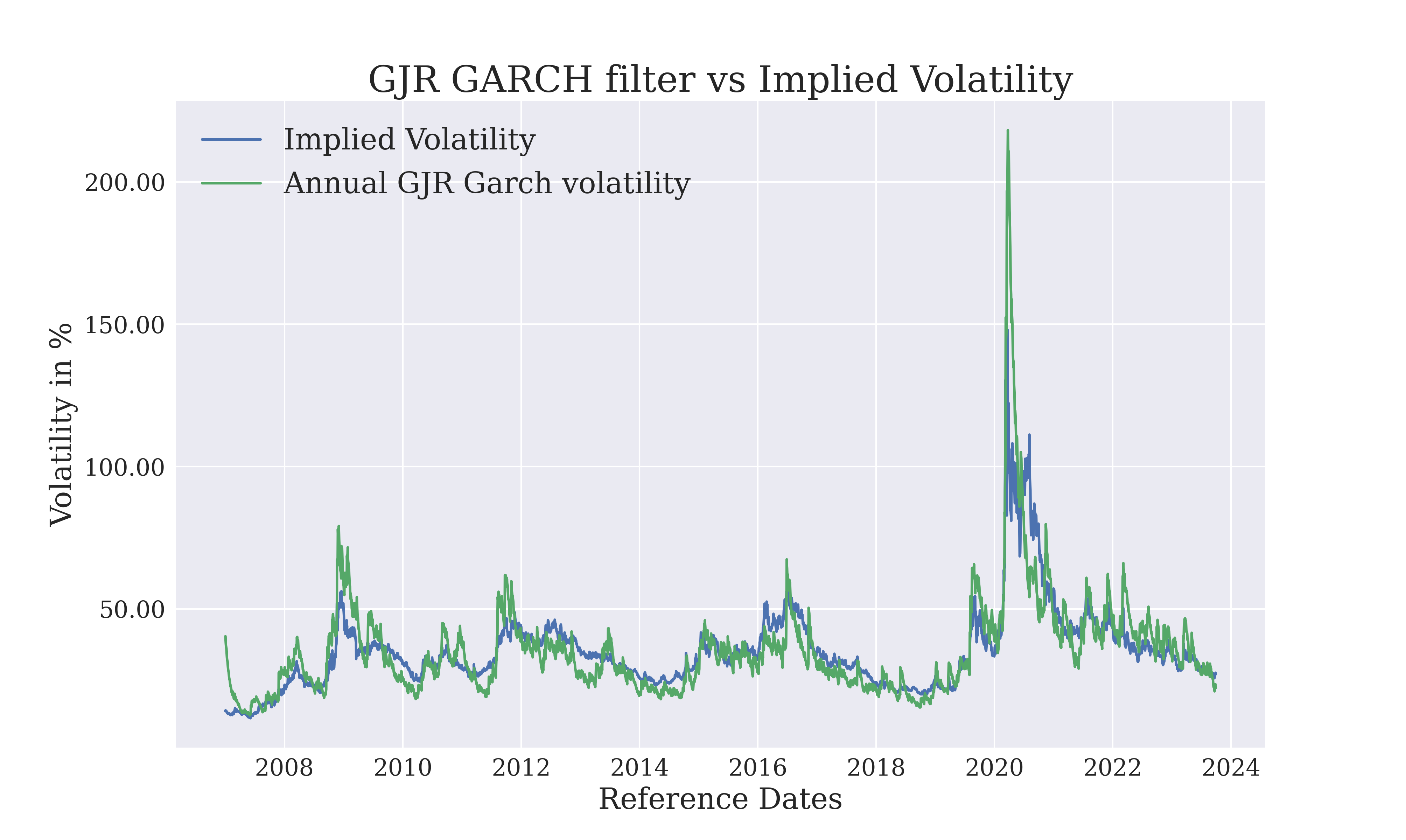}
    \caption{GJR-GARCH model volatility filter. Blue line denotes market implied swaption volatilities and green GJR-GARCH annualized volatility.}
    \label{fig:GJR-GARCH}
\end{figure}

As it can be seen in Figure \ref{fig:GJR-GARCH} and similarly to the standard GARCH case, the GJR-GARCH volatility filter follows
ATM implied volatility closely. Differences between the two models are
quantitatively small, though the asymmetric specification provides a slightly
better match in periods of elevated volatility.

\subsection{Hypotheses testing on model inputs and outputs}
\label{sec:hypotest}

We next investigate the autocorrelation properties of residuals,
squared residuals and standardised residuals, where residuals are understood as a difference between realized returns and their mean. Ljung--Box tests are used to
assess the null hypothesis of no autocorrelation up to a given lag.

Table~\ref{tab:LJUNG-BOX_RES} reports Ljung--Box statistics for raw residuals.

\begin{table}[!ht]
\caption{Ljung-Box testing for residuals (returns - mean) autocorrelation, where header line descriptions stand for test decision, p-value, test statistic, critical value, number of lags, confidence level and degrees of freedom respectively.}
 \label{tab:LJUNG-BOX_RES}
    \centering
\begin{tabular}{|c|c|c|c|c|c|c|c|} \hline
 Test number &  h &   p-value &    stat &  cValue &  Lags &  Alpha &  DoF \\ \hline
           1 &  1 & $<0.0001$ & 19.3152 &  8.1887 &     2 & 0.0167 &    2 \\ \hline
           2 &  1 & $<0.0001$ & 26.2807 & 10.2355 &     3 & 0.0167 &    3 \\ \hline
           3 &  1 & $<0.0001$ & 30.3795 & 13.8388 &     5 & 0.0167 &    5 \\ \hline
\end{tabular}
\end{table}
\FloatBarrier

The null of no autocorrelation in residuals is clearly rejected, which is
consistent with the presence of conditional heteroscedasticity. For squared
residuals, Table~\ref{tab:LJUNG-BOX_SQRRES} confirms even stronger serial
dependence. Both tests confirm the strong ARCH effect already identified earlier.
Normality tests for residuals (Table~\ref{tab::Normality}) strongly reject the
Gaussian assumption, indicating heavy tails.

{
\begin{table}[!ht]
\caption{Ljung-Box testing for autocorrelation of squared residuals, where header line descriptions stand for test decision, p-value, test statistic, critical value, number of lags, confidence level and degrees of freedom respectively.}
 \label{tab:LJUNG-BOX_SQRRES}
    \centering
\begin{tabular}{|c|c|c|c|c|c|c|c|} \hline
 Test number &  h &  p-value &   stat &  cValue &  Lags &  Alpha &  DoF \\ \hline
           1 &  1 &       0 & 1460.3 &  8.1887 &     2 & 0.0167 &    2 \\ \hline
           2 &  1 &       0 & 2066.6 & 10.2355 &     3 & 0.0167 &    3 \\  \hline
           3 &  1 &       0 & 4625.8 & 13.8388 &     5 & 0.0167 &    5 \\   \hline
\end{tabular}
\end{table}

\begin{table}[!ht]
\caption{Tests for residuals' normality, h denotes opted hypothesis, 0 stands for null and 1 for alternative, 
Statistic stands for test statistic and p-value denotes test p-value.}
 \label{tab::Normality}
    \centering
\begin{tabular}{|c|c|c|c|} \hline
           Test name &  h &  Statistic &        p-value \\ \hline
        Shapiro-Wilk &  1 &    0.908665 &  $<0.001$ \\ \hline
D'Agostino-Pearson &  1 &  943.084187 & $<0.001$ \\ \hline
  Kolmogorov-Smirnov &  1 &    0.464300 &  $<0.001$ \\ \hline
\end{tabular}
\end{table}
\FloatBarrier

After filtering with GARCH, GJR-GARCH and MSGARCH Engle’s ARCH test applied to
standardised residuals (Table~\ref{tab::ARCH_STD}) no longer detects
conditional heteroscedasticity, suggesting that the models have successfully
captured the volatility clustering in the data. Ljung--Box tests on standardised residuals as presented in Tables~\ref{tab::LJUNG-BOX-std_RES_GARCH},
~\ref{tab::LJUNG-BOX-std_RES_GJR} and ~\ref{tab::LJUNG-BOX-std_RES_MS} indicate that there is no remaining serial
autocorrelation for either specification, which speaks in favour of correct
model specification.

\begin{table}[!ht]
\caption{Engle's ARCH test for standardized residuals for ATMF swap rates}
 \label{tab::ARCH_STD}
    \centering
\begin{tabular}{|c|c|c|} \hline
    Model &  h &  p-value \\ \hline
    GARCH &  0 & 0.47530 \\ \hline
GJR-GARCH &  0 & 0.53309 \\ \hline
MSGARCH   &  0 &  0.3410  \\ \hline
\end{tabular}
\end{table}
\FloatBarrier

\begin{table}[!ht]
\caption{Ljung-Box test for standardized residuals -- GARCH model, where header line descriptions stand for test decision, p-value, test statistic, critical value, number of lags, confidence level and degrees of freedom respectively}
 \label{tab::LJUNG-BOX-std_RES_GARCH}
    \centering
\begin{tabular}{|c|c|c|c|c|c|c|c|} \hline
 Test number &  h &  p-value &   stat &  cValue &  Lags &  Alpha &  DoF \\ \hline
           1 &  0 &  0.9427 & 0.1179 &  8.1887 &     2 & 0.0167 &    2 \\ \hline
           2 &  0 &  0.9842 & 0.1574 & 10.2355 &     3 & 0.0167 &    3 \\ \hline
           3 &  0 &  0.6061 & 3.6150 & 13.8388 &     5 & 0.0167 &    5 \\ \hline
\end{tabular}
\end{table}

\begin{table}[!ht]
\caption{Ljung-Box test for standardized residuals -- GJR-GARCH model, where header line descriptions stand for test decision, p-value, test statistic, critical value, number of lags, confidence level and degrees of freedom respectively }
 \label{tab::LJUNG-BOX-std_RES_GJR}
    \centering
\begin{tabular}{|c|c|c|c|c|c|c|c|} \hline
 Test number &  h &  p-value &   stat &  cValue &  Lags &  Alpha &  DoF \\ \hline
           1 &  0 &  0.9367 & 0.1309 &  8.1887 &     2 & 0.0167 &    2 \\ \hline
           2 &  0 &  0.9798 & 0.1861 & 10.2355 &     3 & 0.0167 &    3 \\  \hline
           3 &  0 &  0.6189 & 3.5296 & 13.8388 &     5 & 0.0167 &    5 \\   \hline
\end{tabular}
\end{table}
\FloatBarrier

\begin{table}[!ht]
\caption{Ljung-Box test for standardized residuals -- MSGARCH model, where header line descriptions stand for test decision, p-value, test statistic, critical value, number of lags, confidence level and degrees of freedom respectively }
 \label{tab::LJUNG-BOX-std_RES_MS}
    \centering
\begin{tabular}{|c|c|c|c|c|c|c|c|} \hline
 Test number &  h &  p-value &   stat &  cValue &  Lags &  Alpha &  DoF \\ \hline
           1 &  0 &  0.9647 & 0.0719 &  8.1887 &     2 & 0.0167 &    2 \\ \hline
           2 &  0 &  0.9888 & 0.1240 & 10.2355 &     3 & 0.0167 &    3 \\  \hline
           3 &  0 &  0.5962 & 3.6810 & 13.8388 &     5 & 0.0167 &    5 \\   \hline
\end{tabular}
\end{table}
\FloatBarrier

}

{
\subsection{Advanced diagnostic tests}
\label{sec:diagnostics}

Beyond basic residual diagnostics, we employ three sets of advanced tests:
(i) parameter stability (Nyblom test \cite{Nyblom}), (ii) asymmetry diagnostics (Engle--Ng
sign and size bias tests \cite{EngleNg}), and (iii) goodness-of-fit checks for alternative
innovation distributions.

\subsubsection{Parameter stability: Nyblom test}
\label{sec:nyblom}

We examine the temporal stability of parameters in several conditional
volatility models applied to the log returns of the swaption-implied volatility
series. Given the well-documented structural breaks in interest rate markets
--- including the 2008 financial crisis, the prolonged low-volatility
environment induced by quantitative easing (2011--2021), and the volatility
spike in 2022 --- it is essential to assess whether standard GARCH-type models
provide stable and reliable parameter estimates.

Four model specifications are evaluated:
\begin{enumerate}
    \item GARCH(1,1) with normal innovations,
    \item GJR-GARCH(1,1) with normal innovations,
    \item GARCH(1,1) with Student-\(t\) innovations,
    \item GJR-GARCH(1,1) with Student-\(t\) innovations.
\end{enumerate}

Parameter stability is assessed using the Nyblom (1989) test, which evaluates
both individual and joint constancy of model parameters over time. Table \ref{tab:nyblom_garch_norm} presents 
results for regular GARCH model with normal innovations.

\begin{table}[ht!]
    \centering
    \caption{Nyblom test for GARCH(1,1) with normal innovations.
    Columns show: test statistic and critical values at 10\%, 5\%, and 1\% significance levels.
    The row labelled \emph{Joint} reports the Nyblom statistic for the joint null
hypothesis that all model parameters are constant over time.}
    \begin{tabular}{lrrrr}
        \toprule
        Parameter & Statistic & 10\% crit & 5\% crit & 1\% crit \\ \midrule
        $\mu$      & 0.1306 & 0.353 & 0.470 & 0.748 \\
        $\omega$   & 6.6249 & 0.353 & 0.470 & 0.748 \\
        $\alpha_1$ & 0.0726 & 0.353 & 0.470 & 0.748 \\
        $\beta_1$  & 0.1469 & 0.353 & 0.470 & 0.748 \\ \midrule
        Joint      & 19.5856 & 1.07 & 1.24 & 1.60 \\ \bottomrule
    \end{tabular}
    \label{tab:nyblom_garch_norm}
\end{table}
\FloatBarrier

For the GARCH(1,1) model with normal innovations, the intercept parameter
$\omega$ exhibits severe instability, far exceeding the Nyblom critical
values. The remaining parameters, including $\alpha_1$ and $\beta_1$, remain
individually stable. This indicates that short-run volatility dynamics are
relatively consistent over time, even though the long-run variance level
undergoes significant structural shifts. The joint statistic nevertheless
surpasses all critical thresholds, which leads to a rejection of overall
parameter stability. Table \ref{tab:nyblom_gjrgarch_norm} presents respective results of Nyblom test for GJR-GARCH(1,1).
 
\begin{table}[ht!]
    \centering
    \caption{Nyblom test for GJR-GARCH(1,1) with normal innovations. Columns show: test statistic and critical values at 10\%, 5\%, and 1\% significance levels.
    The row labelled \emph{Joint} reports the Nyblom statistic for the joint null
hypothesis that all model parameters are constant over time.}
    \begin{tabular}{lrrrr}
        \toprule
        Parameter & Statistic & 10\% crit & 5\% crit & 1\% crit \\ \midrule
        $\mu$      & 0.1501 & 0.353 & 0.470 & 0.748 \\
        $\omega$   & 10.3567 & 0.353 & 0.470 & 0.748 \\
        $\alpha_1$ & 0.0502 & 0.353 & 0.470 & 0.748 \\
        $\beta_1$  & 0.1397 & 0.353 & 0.470 & 0.748 \\
        $\gamma_1$ & 0.0533 & 0.353 & 0.470 & 0.748 \\ \midrule
        Joint      & 33.0888 & 1.28 & 1.47 & 1.88 \\ \bottomrule
    \end{tabular}
    \label{tab:nyblom_gjrgarch_norm}
\end{table}
\FloatBarrier

Introducing asymmetry through the GJR specification yields stable estimates for
the parameters $\alpha_1$, $\beta_1$, $\mu$ and the leverage parameter
$\gamma_1$. However, the intercept $\omega$ remains highly unstable and the
joint test again rejects stability, indicating that long-run volatility levels
change over time. Table \ref{tab:nyblom_garch_std} presents respective results of Nyblom test for GARCH(1,1).

\begin{table}[ht!]
    \centering
    \caption{Nyblom test for GARCH(1,1) with Student-$t$ innovations. Columns show: test statistic and critical values at 10\%, 5\%, and 1\% significance levels.
    The row labelled \emph{Joint} reports the Nyblom statistic for the joint null
hypothesis that all model parameters are constant over time.}
    \begin{tabular}{lrrrr}
        \toprule
        Parameter & Statistic & 10\% crit & 5\% crit & 1\% crit \\ \midrule
        $\mu$      & 0.3213 & 0.353 & 0.470 & 0.748 \\
        $\omega$   & 12.6071 & 0.353 & 0.470 & 0.748 \\
        $\alpha_1$ & 0.0606 & 0.353 & 0.470 & 0.748 \\
        $\beta_1$  & 0.1692 & 0.353 & 0.470 & 0.748 \\
        $\text{shape}$ & 0.2919 & 0.353 & 0.470 & 0.748 \\ \midrule
        Joint      & 39.5233 & 1.28 & 1.47 & 1.88 \\ \bottomrule
    \end{tabular}
    \label{tab:nyblom_garch_std}
\end{table}
\FloatBarrier

Under Student-$t$ innovations, $\alpha_1$, $\beta_1$ and $\mu$ remain individually
stable, but both the long-run variance parameter $\omega$ and the tail
parameter \emph{shape} display strong instability. The joint statistic again
rejects the null of constant parameters. Table \ref{tab:nyblom_gjrgarch_std} presents results for GJR-GARCH model with t-Student innovations.  

\begin{table}[ht!]
    \centering
    \caption{Nyblom test for GJR-GARCH(1,1) with Student-$t$ innovations. Columns show: test statistic and critical values at 10\%, 5\%, and 1\% significance levels.
    The row labelled \emph{Joint} reports the Nyblom statistic for the joint null
hypothesis that all model parameters are constant over time.}
    \begin{tabular}{lrrrr}
        \toprule
        Parameter & Statistic & 10\% crit & 5\% crit & 1\% crit \\ \midrule
        $\mu$      & 0.3552 & 0.353 & 0.470 & 0.748 \\
        $\omega$   & 15.9059 & 0.353 & 0.470 & 0.748 \\
        $\alpha_1$ & 0.0728 & 0.353 & 0.470 & 0.748 \\
        $\beta_1$  & 0.1560 & 0.353 & 0.470 & 0.748 \\
        $\gamma_1$ & 0.1086 & 0.353 & 0.470 & 0.748 \\
        $\text{shape}$ & 0.2895 & 0.353 & 0.470 & 0.748 \\ \midrule
        Joint      & 57.3712 & 1.49 & 1.68 & 2.12 \\ \bottomrule
    \end{tabular}
    \label{tab:nyblom_gjrgarch_std}
\end{table}
\FloatBarrier

The GJR-GARCH model with Student-$t$ innovations exhibits stable
parameters ($\alpha_1$, $\beta_1$, $\gamma_1$, $\mu$), but strong instability in
$\omega$ and the \emph{shape} parameter. The joint statistic decisively rejects
overall stability.

\medskip\noindent
Across all four specifications, short-run volatility parameters governing
clustering and leverage effects remain individually stable, whereas the
intercept $\omega$ and, in heavy-tailed models, the tail parameter are clearly
time-varying. The Nyblom joint tests strongly reject constant-parameter
specifications, pointing towards volatility regime shifts that single-regime
GARCH models cannot accommodate. This motivates the regime-switching analysis
developed in Section~\ref{sec:msgarch}.

\subsubsection{Goodness-of-fit of innovation distributions: Adjusted Pearson test}
\label{subsec:gof_summary}

We next assess whether the assumed conditional distribution of innovations is
consistent with the empirical behaviour of standardised residuals. Four
candidate families widely used in the volatility literature are considered:
Gaussian, Student-\emph{t}, GED \cite{Subbotin1923}, \cite{Nelson1991} and Johnson~SU \cite{JohnsonSU}. Model adequacy is evaluated
using the adjusted Pearson chi-squared goodness-of-fit statistic based on
quantile-balanced bins. Results are summarised in
Table~\ref{tab:pearson_all}.

\begin{table}[ht!]
\centering
\caption{Adjusted Pearson goodness-of-fit test for alternative innovation distributions.}
\label{tab:pearson_all}
\begin{tabular}{lccc}
\toprule
Distribution & $\chi^{2}$ statistic & $p$-value & Interpretation \\ \midrule
Normal            & 102.17  & $2.2 \times 10^{-13}$ & Rejected \\
Student-\emph{t}  & 4364.00 & 0.00                   & Strongly rejected \\
GED               & 133.87  & 0.00                   & Rejected \\
Johnson~SU        & 36.56   & 0.00147                & Best fit; marginally rejected \\ \bottomrule
\end{tabular}
\end{table}
\FloatBarrier

The Gaussian distribution is clearly inadequate, as indicated by a highly 
significant Pearson statistic. Student-\emph{t} improves the maximised 
log-likelihood during estimation, but its goodness-of-fit performance is even 
worse than under the Gaussian benchmark, reflecting excessively heavy tails 
relative to the empirical distribution. The GED distribution also fails to 
improve the Pearson statistic and performs slightly worse than the Gaussian 
case. By contrast, the Johnson~SU family provides by far the best overall fit, 
capturing both skewness and heavy tails, although the null hypothesis is still 
rejected at conventional levels. In practice we adopt Normal or the Student-\emph{t} 
distribution as the baseline, while treating Johnson~SU as a robustness benchmark in the 
diagnostic analysis.

\subsubsection{Asymmetry diagnostics: Engle--Ng sign bias tests}
\label{subsec:engle_ng}

To evaluate whether the conditional variance specification adequately captures
the asymmetric response of volatility to past shocks, we employ the Engle--Ng
(1993) sign bias, negative size bias and positive size bias tests. These
regressions examine whether squared standardised residuals react differently to
positive and negative innovations not explicitly modelled by the variance
equation. Table~\ref{tab:engle_ng} reports the test statistics for both the
symmetric GARCH(1,1) and asymmetric GJR-GARCH(1,1) specifications.

\begin{table}[ht!]
\centering
\caption{Engle--Ng sign, negative size, and positive size bias tests.}
\label{tab:engle_ng}
\begin{tabular}{lcccccc}
\toprule
& \multicolumn{3}{c}{GARCH(1,1)} & \multicolumn{3}{c}{GJR--GARCH(1,1)} \\
\cmidrule(lr){2-4} \cmidrule(lr){5-7}
Test & $t$-stat & $p$-value & Joint $p$ &
       $t$-stat & $p$-value & Joint $p$ \\
\midrule
Sign bias            &  0.956 & 0.339 & --      
                     & \textit{n/a} & \textit{n/a} & --    \\
Negative size bias   & -0.866 & 0.386 & --      
                     & \textit{n/a} & \textit{n/a} & --    \\
Positive size bias   & -0.708 & 0.479 & --      
                     & -0.063 & 0.950 & --        \\
Joint test           &  --    &  --   & 0.794   
                     & --     &  --   & 0.950     \\
\bottomrule
\end{tabular}

\vspace{0.8em}
\begin{flushleft}
\footnotesize
\textbf{Notes:} The Engle--Ng sign and negative size bias tests are not applicable 
for the GJR--GARCH model because asymmetry is already explicitly built into the 
conditional variance equation via the leverage term $\gamma$. This induces 
perfect collinearity between the Engle--Ng regressors and the model-implied 
indicator for negative shocks, preventing coefficient identification. Hence the 
statistics are reported as ``n/a''. The row labelled \emph{Joint test} reports the 
$p$-values of the joint Engle--Ng test for all three bias terms.
\end{flushleft}
\end{table}
\FloatBarrier

For the symmetric GARCH(1,1) model, individual sign and size effects are not
significant, but $p$-values around 0.34--0.39 indicate mild residual asymmetry
that is not explicitly captured by the variance equation. The joint Engle--Ng
statistic yields a $p$-value of 0.79, suggesting that asymmetry is not
statistically pronounced but remains informative about the limits of the
symmetric specification.

For the GJR-GARCH(1,1) model, all individual sign and size bias coefficients
are insignificant and the joint test yields a $p$-value of 0.95. Once the
leverage term is introduced directly in the conditional variance equation, no
residual asymmetric effects remain in the standardised residuals. This confirms
that the GJR-GARCH specification provides a more appropriate representation of
volatility dynamics in the swap-rate data.

Summirizing, Standard GARCH and GJR-GARCH models with Student-\emph{t} innovations successfully remove
autocorrelation and conditional heteroscedasticity from standardised residuals,
and the GJR specification effectively captures the modest asymmetry present in
the data. However, Nyblom tests reveal pronounced instability in the
intercept and, for heavy-tailed models, in the tail parameter, indicating that
long-run volatility levels and tail behaviour change over time. Goodness-of-fit
tests further show that Gaussian innovations are inadequate and that more
flexible distributions are required.

Overall, these diagnostics suggest that while single-regime GARCH-type models
provide a reasonable short-run description of volatility clustering and
asymmetry, they are unable to accommodate the structural regime shifts present
in the 2007--2023 sample. This motivates the regime-switching framework
developed in Section~\ref{sec:msgarch}, where MSGARCH models
with heavy-tailed innovations are introduced to capture time-varying volatility
states in a coherent way.
}

{ 

\section{Application of Regime-switching GARCH with heavy-tailed innovations}
\label{sec:msgarch}

The diagnostic analysis performed in Section~\ref{sec:diagnostics} revealed
two important features of the data. 
First, the Nyblom stability test strongly rejects the null of constant
GARCH parameters over the 2007--2023 period, indicating the presence of 
structural breaks and changes in volatility dynamics. 
Second, goodness-of-fit tests show that fat-tailed distributions 
provide a markedly better fit to standardized residuals than the Gaussian
specification. 
Together, these two findings motivate the use of a 
MSGARCH model with heavy-tailed innovations.

\subsection{Estimation results: Student-\texorpdfstring{$t$}{t} innovations}

Table \ref{tab:msgarch-t} reports the MLE results for the 
MSGARCH(1,1) model with Student-$t$ innovations.

\begin{table}[ht!]
\centering
\caption{Parameter estimates for the two-regime MSGARCH(1,1) model with Student-$t$ innovations.}
\label{tab:msgarch-t}
\begin{tabular}{|c|c|c|}
\hline
Parameter & Regime 1 & Regime 2 \\ \hline
$\mu$     & \multicolumn{2}{c|}{$-0.00017$} \\ \hline
$\omega$  & $9.72 \times 10^{-5}$ & $4.11 \times 10^{-7}$ \\ \hline
$\alpha$  & $0.352$ & $0.033$ \\ \hline
$\beta$   & $0.599$ & $0.958$ \\ \hline
$\nu$ (degrees of freedom) & \multicolumn{2}{c|}{$14.59$} \\ \hline
$p_{11}$ & \multicolumn{2}{c|}{$0.450$} \\ \hline
$p_{22}$ & \multicolumn{2}{c|}{$0.921$} \\ \hline
$p_{12}$ & \multicolumn{2}{c|}{$0.550$} \\ \hline
$p_{21}$ & \multicolumn{2}{c|}{$0.079$} \\ \hline
Log-likelihood & \multicolumn{2}{c|}{$-10771.29$} \\ \hline
\end{tabular}
\end{table}
\FloatBarrier

The model converges successfully (i.e.\ the maximum-likelihood optimiser reaches 
a stable parameter vector satisfying the first-order conditions) with a negative 
log-likelihood of $-10{,}771.29$, and produces well-separated regimes.
Regime~1 exhibits lower GARCH persistence ($\alpha_1=0.3518$, $\beta_1=0.5987$),
while Regime~2 shows near-unit persistence ($\beta_2=0.9576$), consistent with
high-volatility episodes.
The transition probabilities,
\[
P = 
\begin{pmatrix}
0.4496 & 0.5504 \\
0.0794 & 0.9206
\end{pmatrix},
\]
indicate a highly persistent high-volatility regime.
The degrees of freedom parameter ($\nu = 14.59$) confirms the presence of
moderate fat tails.

Figure~\ref{fig:msgarch-t} compares the mixed MSGARCH volatility with the
ATM swaption implied volatility. 

\begin{figure}[ht!]
    \centering
    \includegraphics[width=1.0\linewidth]{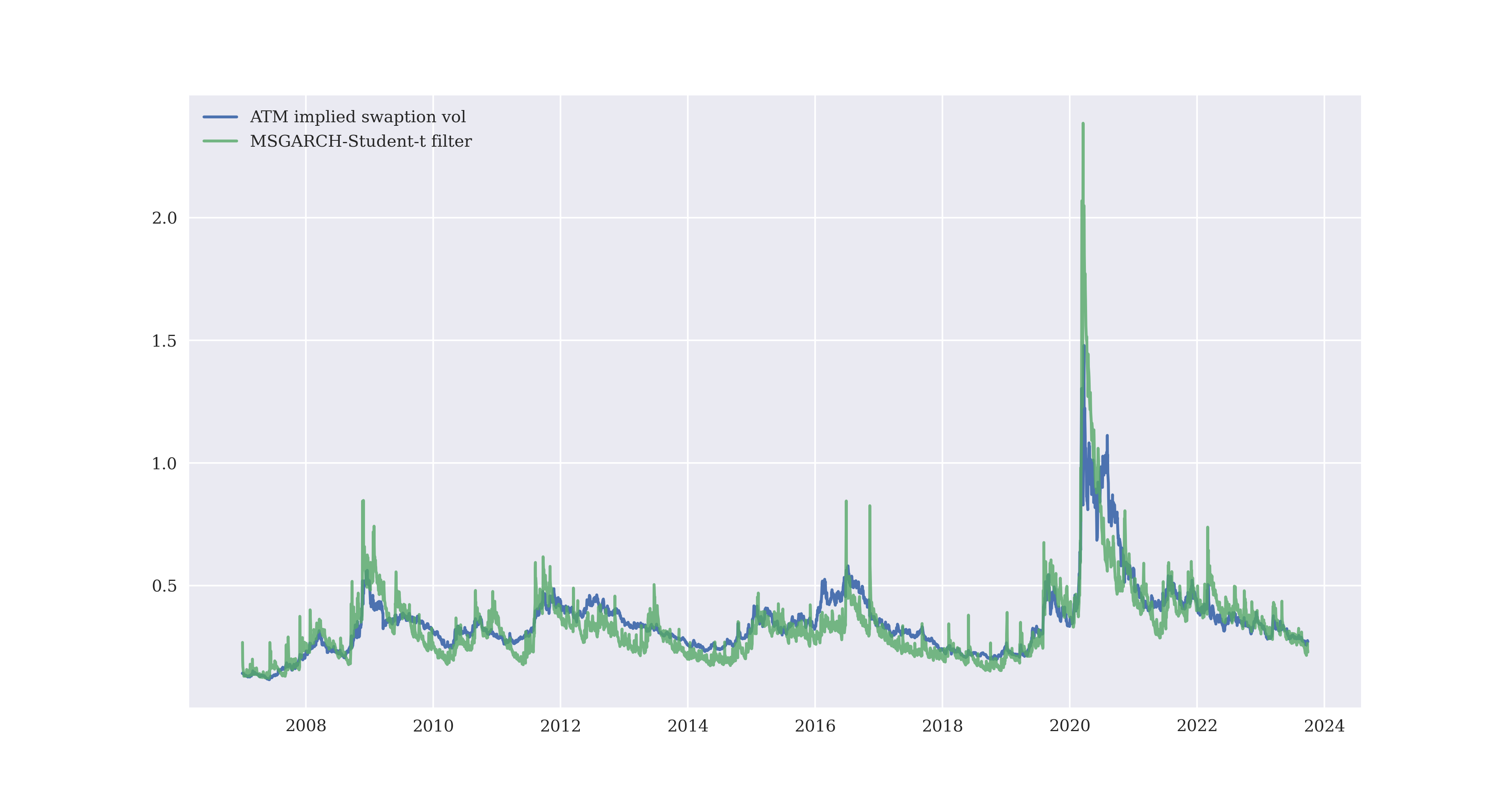}
    \caption{MSGARCH-Student t versus ATM implied vol. Blue line denotes historical ATM Swaption volatilities, whereas green MSGARCH filtration}
    \label{fig:msgarch-t}
\end{figure}

\FloatBarrier

The filtered path captures both the rapid volatility spikes (e.g., 2008,
2011--2012, 2020) and the long periods of low volatility, with a close visual
alignment to market-implied measures.

\subsection{Estimation results: Johnson~\texorpdfstring{$SU$}{SU} innovations}

As a robustness check we also estimate the MSGARCH(1,1) specification using Johnson~$SU$ innovations, calibrated to 
standardized residuals from a preliminary GJR-GARCH model fitted to 
appropriately rescaled returns.
The resulting log-likelihood equals -$9{,}398.62$ and the optimizer converges
successfully. The estimated parameters (see Table~\ref{tab:msgarch_jsu})
again produce two well-defined regimes with persistence patterns similar to
the Student-$t$ specification.

\begin{table}[ht!]
\centering
\caption{Parameter estimates for the two-regime MSGARCH(1,1) model with Johnson~$SU$ innovations.}
\label{tab:msgarch_jsu}
\begin{tabular}{|c|c|c|}
\hline
Parameter & Regime 1 & Regime 2 \\ \hline
$\mu$     & \multicolumn{2}{c|}{-0.03} \\ \hline
$\omega$  & 0.17 & 0.001 \\ \hline
$\alpha$  & 0.14 & 0.03 \\ \hline
$\beta$   & 0.85 & 0.97 \\ \hline
\multicolumn{3}{|c|}{\textbf{Johnson $SU$ innovation parameters}} \\ \hline
Shape $\gamma$ & \multicolumn{2}{c|}{0.002} \\ \hline
Shape $\delta$ & \multicolumn{2}{c|}{1.28} \\ \hline
Location $\xi$ & \multicolumn{2}{c|}{0.0} \\ \hline
Scale $\lambda$ & \multicolumn{2}{c|}{1.0} \\ \hline
\multicolumn{3}{|c|}{\textbf{Transition probabilities}} \\ \hline
$p_{11}$ & \multicolumn{2}{c|}{0.75} \\ \hline
$p_{22}$ & \multicolumn{2}{c|}{0.86} \\ \hline
$p_{12}$ & \multicolumn{2}{c|}{0.25} \\ \hline
$p_{21}$ & \multicolumn{2}{c|}{0.14} \\ \hline
Log-likelihood & \multicolumn{2}{c|}{$-9398.62$} \\ \hline
\end{tabular}
\end{table}
\FloatBarrier

\begin{figure}[ht!]
    \centering
    \includegraphics[width=1.0\linewidth]{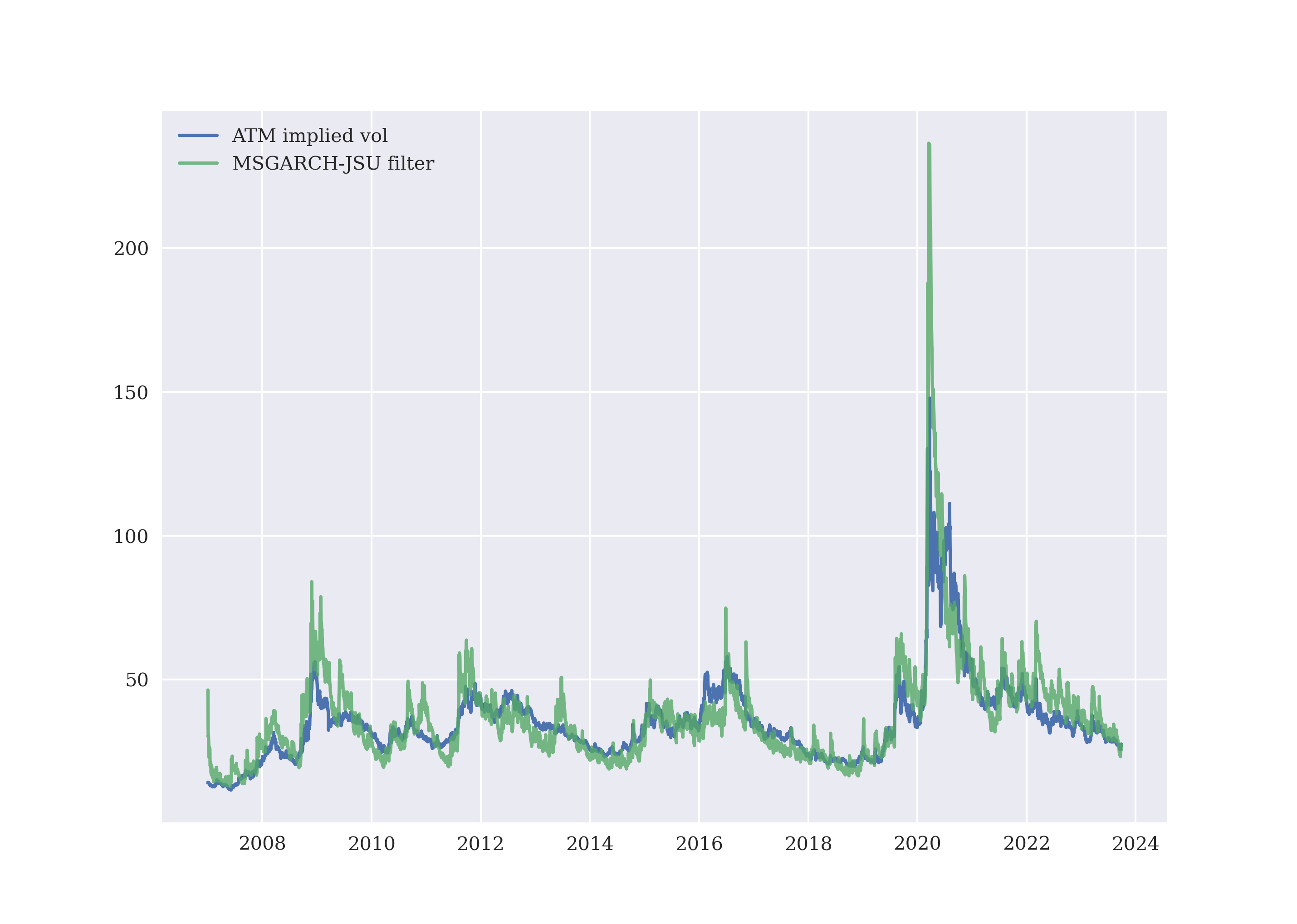}
    \caption{MSGARCH-JSU filter versus ATM implied vol. Blue line denotes historical ATM Swaption volatilities, whereas green MSGARCH filtration}
    \label{fig:msgarch-jsu}
\end{figure}

\FloatBarrier
Figure~\ref{fig:msgarch-jsu} displays the corresponding volatility filter.
While the Johnson~$SU$ distribution provides additional flexibility in skewness
and tail behaviour, the improvement in fit relative to the Student-$t$ 
specification is marginal in the regime-switching setting. 
This suggests that most of the non-Gaussianity is already captured by the 
combination of regime transitions and heavy tails, a finding aligned with the 
literature on MSGARCH models.
We therefore retain the Student-$t$ innovation as the baseline model and view
the Johnson~$SU$ results as an important robustness check.

\subsubsection{Interpretation and relation to parameter instability}

The regime structure identified by the MSGARCH model is consistent with the 
instability detected by the Nyblom test in single-regime GARCH models.
What appears as ``parameter instability'' in a constant-parameter model
manifests naturally as persistent switching between low- and high-volatility
regimes once a Markov-switching structure is allowed.
In particular:
\begin{itemize}
\item the high-persistence regime (Regime 2) corresponds to major market stress events,
including the 2008 financial crisis, the 2013 Federal Reserve tapering episode,
and the volatility surge during the COVID-19 outbreak,
\item the low-persistence regime (Regime 1) prevails during periods of market stability,
characterised by low interest rate levels and persistently subdued volatility,
\item the transition matrix indicates extended durations within each regime,
consistent with the slow structural adjustments associated with the
transitions between LIBOR, OIS and SOFR collateralisation frameworks.
\end{itemize}

% Thus, MSGARCH reconciles the diagnostic evidence: 
% parameter instability in the single-regime model reflects genuine structural 
% segmentation in volatility dynamics rather than model misspecification.
% The regime-switching framework provides a coherent, economically interpretable 
% and statistically stable representation of the observed time series.
A key motivation for introducing the MSGARCH specification arises from the
instability of parameters detected in the single-regime GARCH and GJR-GARCH
models. As shown in Section~\ref{sec:modeltest}, the Nyblom test strongly rejects
the null of constant parameters, indicating that the unconditional variance,
short-run ARCH effect, and persistence components drift over time in a manner
inconsistent with a stationary single-regime structure. This behaviour aligns
with the presence of structural breaks and long-lasting shifts in the interest
rate environment, particularly around the financial crisis, the post-crisis
quantitative easing period, and the transition from LIBOR to OIS and SOFR.

In a regime-switching framework, parameters are not expected to remain constant
throughout the sample; instead, the volatility process is allowed to switch
between distinct, state-dependent dynamics governed by the Markov chain. For
this reason, the classical Nyblom test designed to assess time variation of a
single parameter vector—is not directly applicable to MSGARCH models. Rather
than testing for constancy, regime switching itself provides a structural
representation of the parameter changes detected in the single-regime models.

Accordingly, diagnostic assessment for MSGARCH focuses on evaluating the
coherence and persistence of each regime, the adequacy of regime-specific
innovation distributions, and the ability of filtered volatility to track
market-implied measures such as swaption volatility and SRVIX. These diagnostics
serve as the natural analogue of parameter stability testing in a
multi-regime environment and allow for meaningful comparison with the
single-regime benchmarks.

\subsubsection{Summary}

The MSGARCH analysis addresses adressess the following points:
\begin{enumerate}
\item comparing standard GARCH and GJR-GARCH models with richer alternatives 
      (MSGARCH), thereby extending the modelling framework to account for 
      volatility regime changes;
\item evaluating heavy-tailed and skewed innovation distributions 
      (Student-$t$, Johnson~$SU$);
\item showing that Johnson~$SU$ innovations offer minimal improvements beyond 
      the Student-$t$ specification once regime switching is included;
\item providing an integrated interpretation of parameter instability, regime
      shifts, and market transitions (LIBOR--OIS--SOFR).
\end{enumerate}

Overall, MSGARCH with Student-$t$ innovations delivers acceptable balance of
statistical fit, numerical stability, interpretability, and alignment with 
market-implied volatility. The regime-switching volatility filters obtained from the MSGARCH models are 
broadly consistent with the rolling-window GARCH and GJR-GARCH backtests 
reported in Section~\ref{sec:backtest}. In particular, both approaches
identify the same high-volatility episodes (the 2008 crisis, the 2011--2012
Eurozone stress, the COVID shock) and the same extended low-volatility period
in the mid-2010s, with MSGARCH mainly sharpening the transitions between
regimes. The level and timing of the MSGARCH-based volatility estimates
are also in line with the out-of-sample forecast errors from the backtests,
indicating that allowing for regime switching refines rather than overturns
the conclusions on model performance.

}

\section{Considerations on back-testing}
\label{sec:backtest}
In order to further check upon model calibration quality and specification we performed a back-test, where: 
\begin{enumerate} 
\item We define the set of data to train the model, 
\item Next we define the set of data to test the model,
\item  We iteratively (step-by-step) calibrate the model moving along the time span by one day at a time,
\item With calibrated parameters we calculate a one day variance forecast,
\item Next we set 95\% projection interval defined as $mean \pm 1.96 \cdot \sqrt{\sigma^{2}}$, where $\sigma^2$ is a GARCH projected variance. Where $mean$ stands for the arithmetic average of samples from the training period.
\end{enumerate}
The closer the percentage of realized returns falling into the projection interval is to 95\% the better the model specification.

Table \ref{tab::Back-test results} presents results for the above described back-test procedure.

\begin{table}[H]
\caption{Back-test results for GARCH, GJR-GARCH and MSGARCH where header line description stands for number of test observations, number of training observations, share of test observations falling into projection interval under GJR model, share of test observations falling into projection interval under GARCH model, ratio of test observations to total number of observations.}
 \label{tab::Back-test results}
    \centering
\begin{tabular}{|c|c|c|c|c|c|c|} \hline

\makecell{No.} &  \makecell{Number of \\ test observations} &   \makecell{Number of \\ training observations}  &  \makecell{GJR} &  \makecell{GARCH} &  \makecell{MSGARCH} &  \makecell{\% of total \\ observations} \\ \hline

  1 &                         1500 & 2865 & 94.00\% & 93.80\% &        93.8\% &           34.36\% \\  \hline
  2 &                         1300 & 3065 & 93.62\% & 93.23\% &             93.154\% &      29.78\% \\   \hline
  3 &                         1000 & 3365 & 93.30\% & 93.20\% &              93.0\% &     22.91\% \\   \hline
  4 &                          500 & 3865 & 94.40\% & 94.60\% &              95.0\% &     11.45\% \\    \hline
  5 &                          300 & 4065 & 95.67\% & 95.33\% &                96.0\% &   6.87\% \\     \hline

\end{tabular}

\end{table}
{
The empirical coverage rates for both GARCH(1,1) and GJR-GARCH(1,1) and MSGARCH are close 
to the nominal 95\% level. Differences between the models are minor, consistent 
with the observation that one-step-ahead predictions depend primarily on the most 
recent volatility information and are therefore relatively insensitive to long-term 
structural instabilities. The rolling-window evidence thus complements the 
diagnostic tests: although single-regime GARCH models can deliver reasonable 
short-term forecasts, their parameter instability indicates that they do not provide a 
structurally consistent description of long-run volatility dynamics.
}

As a next step we check how model parameters behave for the back-test calibration we described above. With this, we want to capture changing market sentiment (i.e., higher or lower volatility) and its reflection in the model parameters' values. 

\begin{figure}[H]
    \centering
    \includegraphics[width=1.0\linewidth]{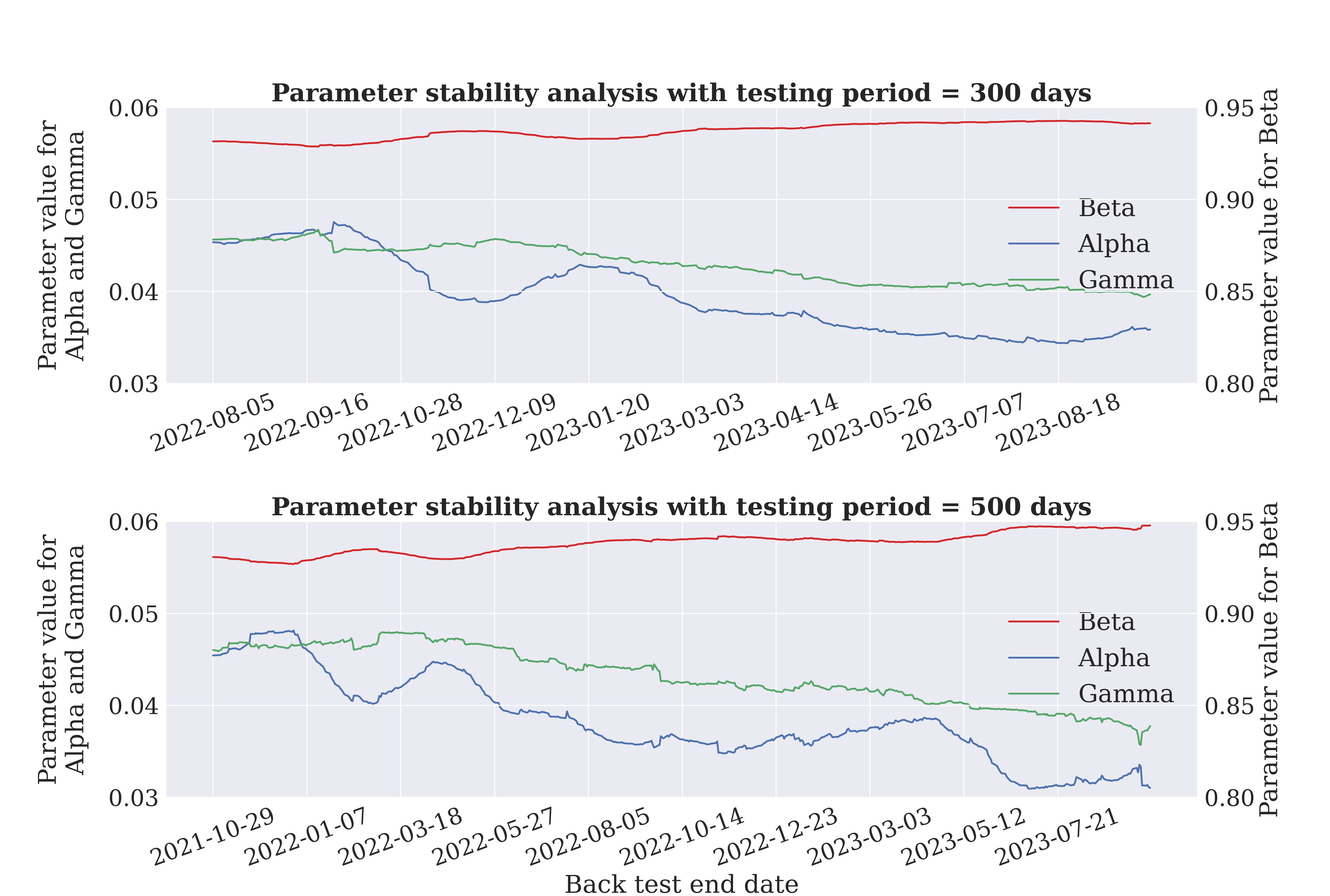}
    \caption{GJR-GARCH parameter stability testing for 300 (top) and 500 (bottom) testing days.}
    \label{fig::Paramater_stability_GJR_1}
\end{figure}

\begin{figure}[H]
    \centering
    \includegraphics[width=1.0\linewidth]{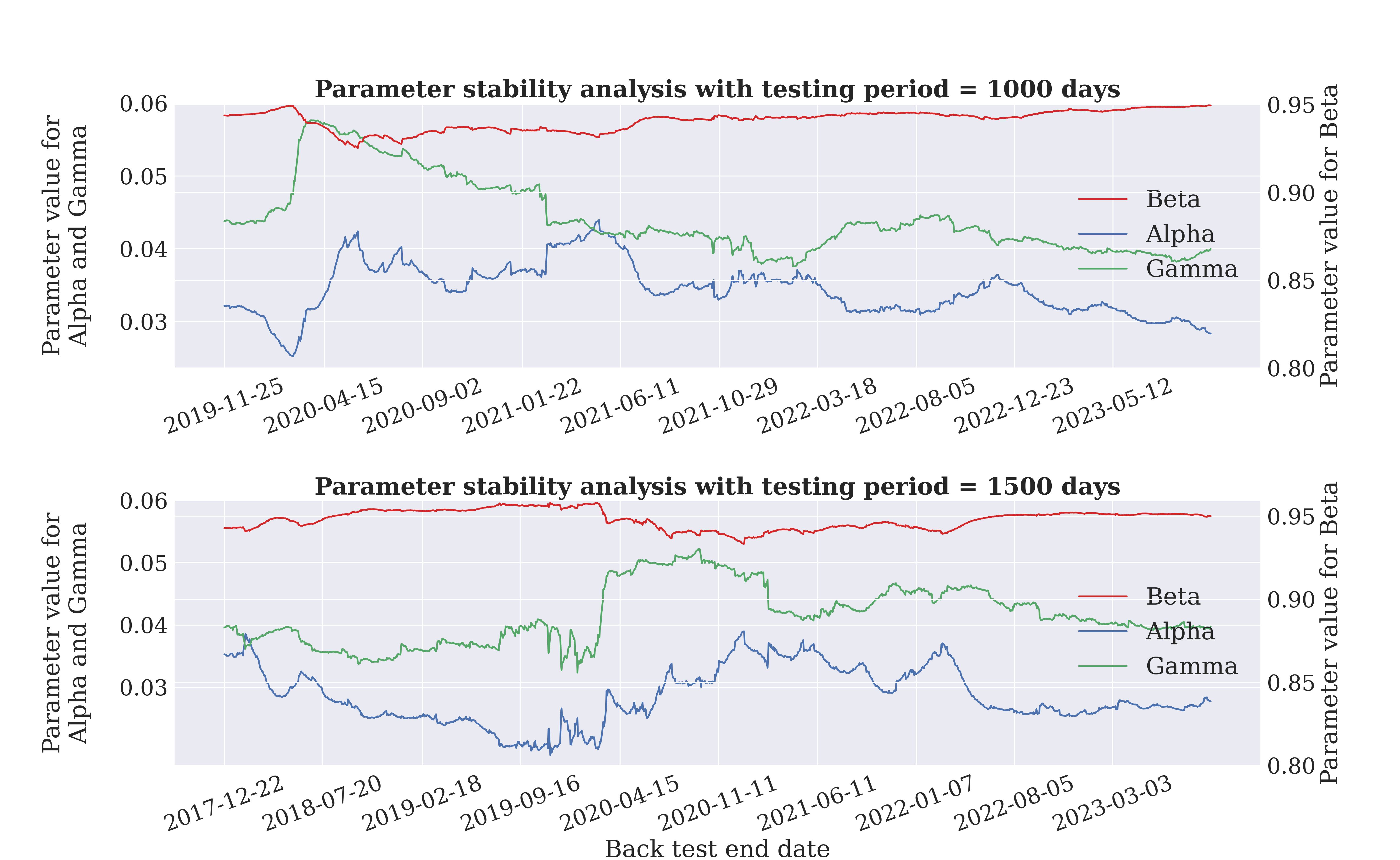}
    \caption{GJR-GARCH parameter stability testing for 1000 (top) and 1500 (bottom) testing days.}
    \label{fig::Paramater_stability_GJR_2}
\end{figure}

Figures \ref{fig::Paramater_stability_GJR_1} and \ref{fig::Paramater_stability_GJR_2} present results for GJR-GARCH model parameter testing along the back-test moving window for a different number of model testing days.

Left hand axis refers to $\alpha$ (ARCH) and $\gamma$ (Leverage) and the right-hand side refers to $\beta$ (GARCH). In general, parameters are stable across the width of the testing period moving $\pm 1\%$ across the line. We can see larger move in $\alpha$ and $\gamma$, i.e. these parameters that measure the contribution of market innovations to the variance levels. For testing periods of 300 and 500 days, these parameters move in line slightly downwards, however for 1000 days testing periods, we can see for that period end falling in the second half of 2020 they diverge with $\alpha$ going up. A similar thing happened in the time range for the period end between March and December 2022. The first is connected to the post-pandemic inflationary environment and the second one to the Ukraine-Russia war outbreak.  More nervousness in the markets dampens volatility persistence and strengthens leverage and ARCH effect, which is reflected in the volatility stylized facts (see section \ref{sec:StylizedFacts})

\begin{figure}[H]
    \centering
    \includegraphics[width=1.0\linewidth]{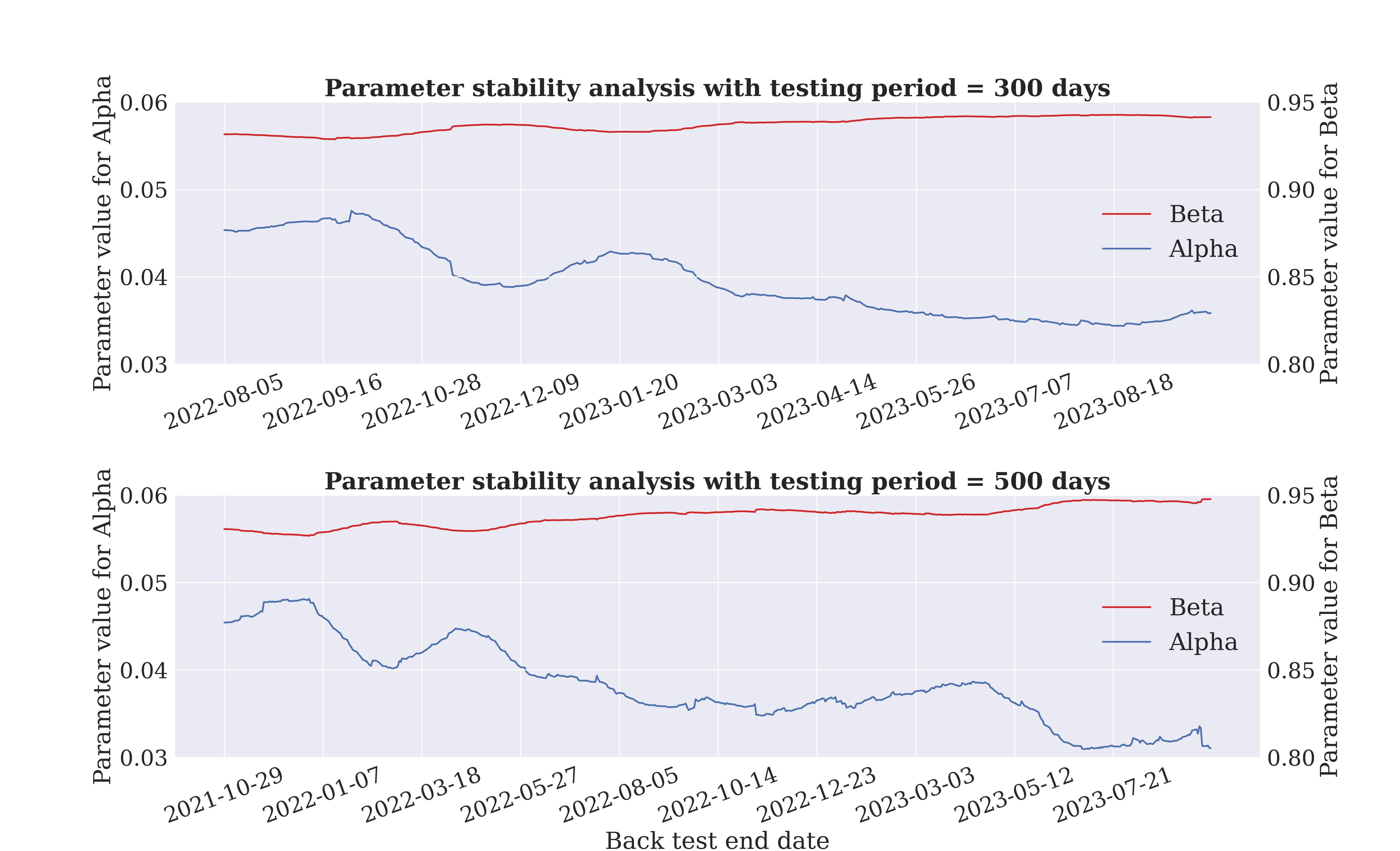}
    \caption{GARCH parameter stability testing for 300 (top) and 500 (bottom) testing days.}
    \label{fig::Paramater_stability_GARCH_1}
\end{figure}

\begin{figure}[H]
    \centering
    \includegraphics[width=1.0\linewidth]{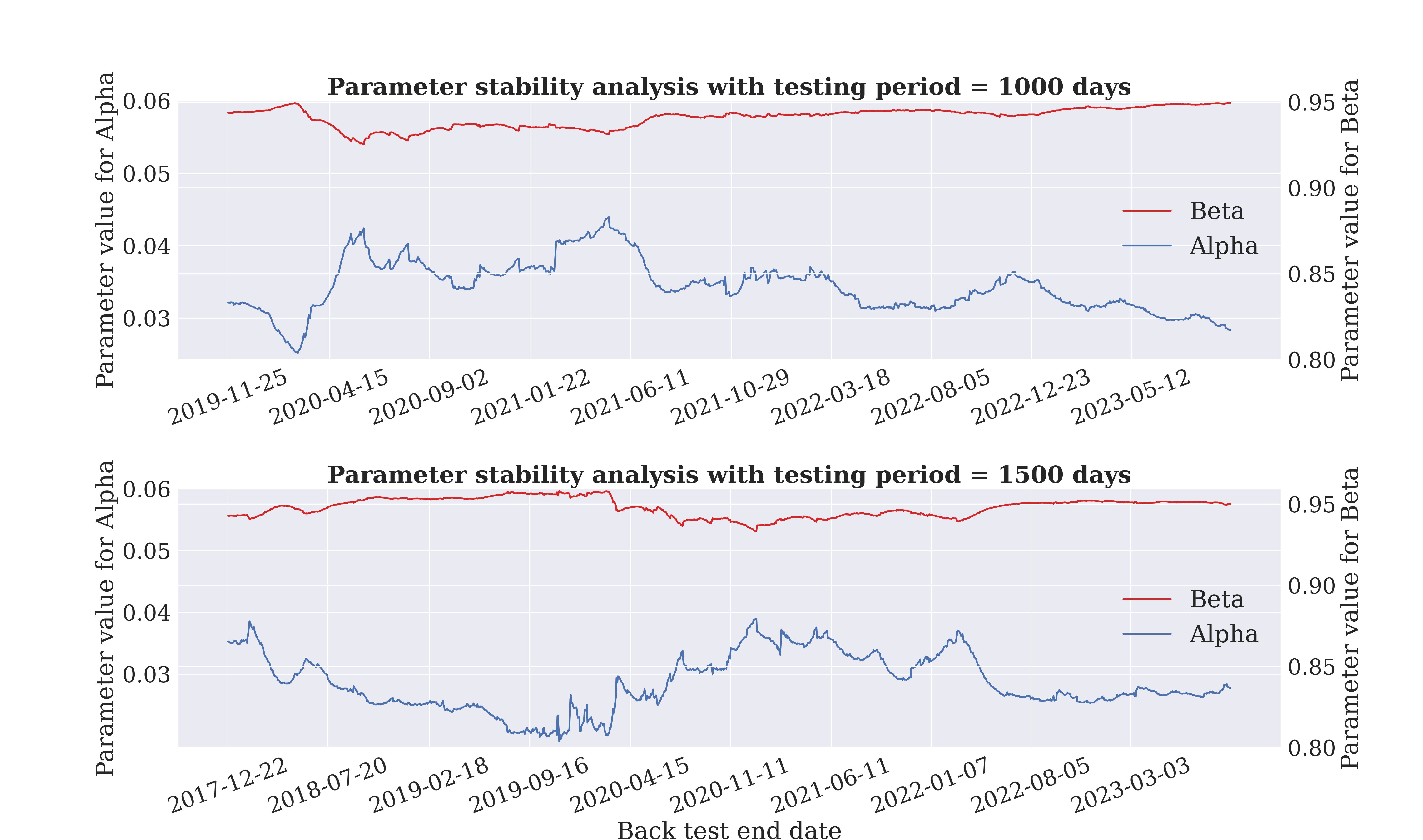}
    \caption{GARCH parameter stability testing for 1000 (top) and 1500 (bottom) testing days.}
    \label{fig::Paramater_stability_GARCH_2}
\end{figure}

Figure \ref{fig::Paramater_stability_GARCH_1} and \ref{fig::Paramater_stability_GARCH_2} depict similar analysis but performed for standard GARCH model.

Top panel shows $\beta$ and $\alpha$ stability analysis for 300 testing days. Both are stable in the examined period with $\alpha$ parameter trending slightly downwards. Similar pattern can be seen in the bottom panel with 500 testing days. Both parameters are stable with $\alpha$ parameter trending downwards, which speaks of less nervousness in the markets during examined period. 

\begin{figure}[H]
    \centering
    \includegraphics[width=1.0\linewidth]{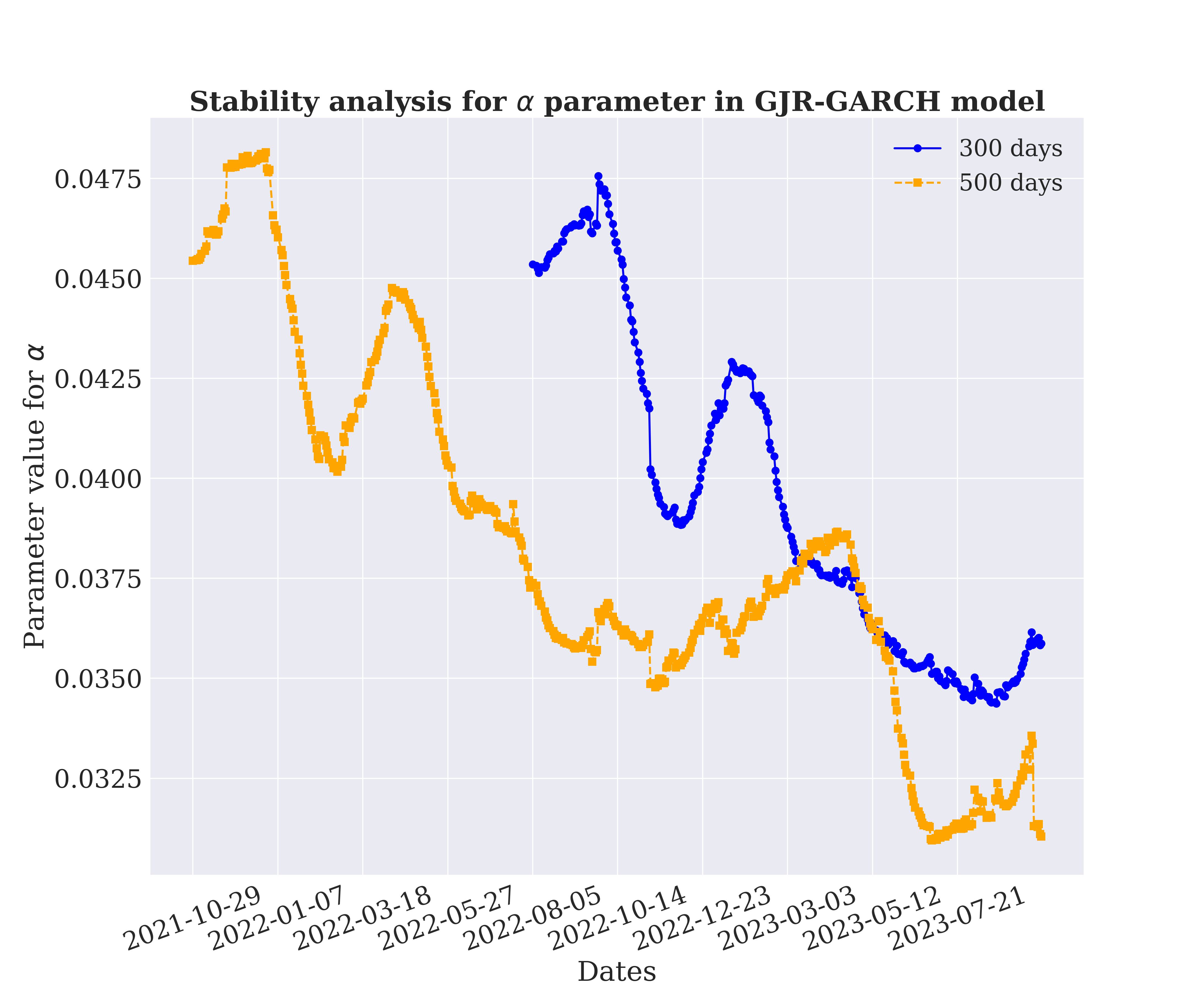}
    \caption{GJR-GARCH $\alpha$ parameter stability testing for short testing periods.}
    \label{fig::Paramater_stability_ALPHA_GJR_2}
\end{figure}

We can see that both parameters are stable. With 1500 testing days we can see a sharp increase around the beginning of 2020. This is the COVID-19 pandemic, showing model response to elevated market nervousness. Taking a closer look into the parameters' behavior we can confirm what we said above. COVID-19 introduced higher volatility to the markets which resulted in increased values of $\alpha$ parameter. A similar conclusion can be drawn for the shorter testing period as presented in Figure \ref{fig::Paramater_stability_ALPHA_GJR_2}.

\begin{figure}[H]
    \centering
    \includegraphics[width=1.0\linewidth]{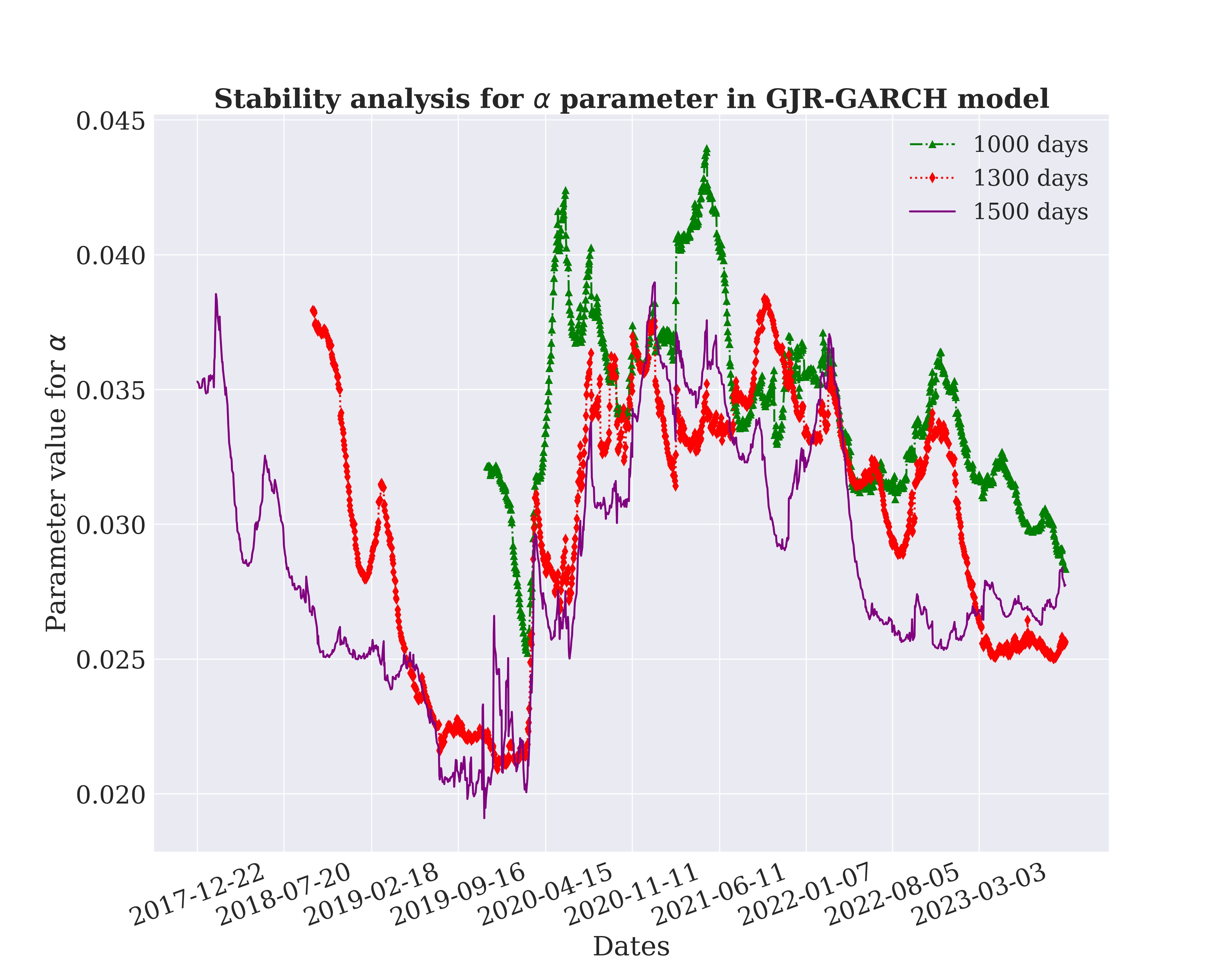}
    \caption{GJR-GARCH $\alpha$ parameter stability testing for long testing periods}
    \label{fig::Paramater_stability_ALPHA_GJR_1}
\end{figure}

Figure \ref{fig::Paramater_stability_ALPHA_GJR_1} presents an analysis for $\alpha$ parameter performed for GJR-GARCH model.

With longer testing periods we can see jump in $\alpha$ parameter values at the beginning of 2020. This is caused by the COVID-19, that brought more nervousness to the financial markets and thus increased volatility.  

Shorter period analysis for $\alpha$ parameter stability starts in late 2021 and finishes in late 2023, we can see that situation stabilizes after COVID-19 and $\alpha$ parameter is trending downwards, which speaks of falling volatility.

\begin{figure}[H]
    \centering
    \includegraphics[width=1.0\linewidth]{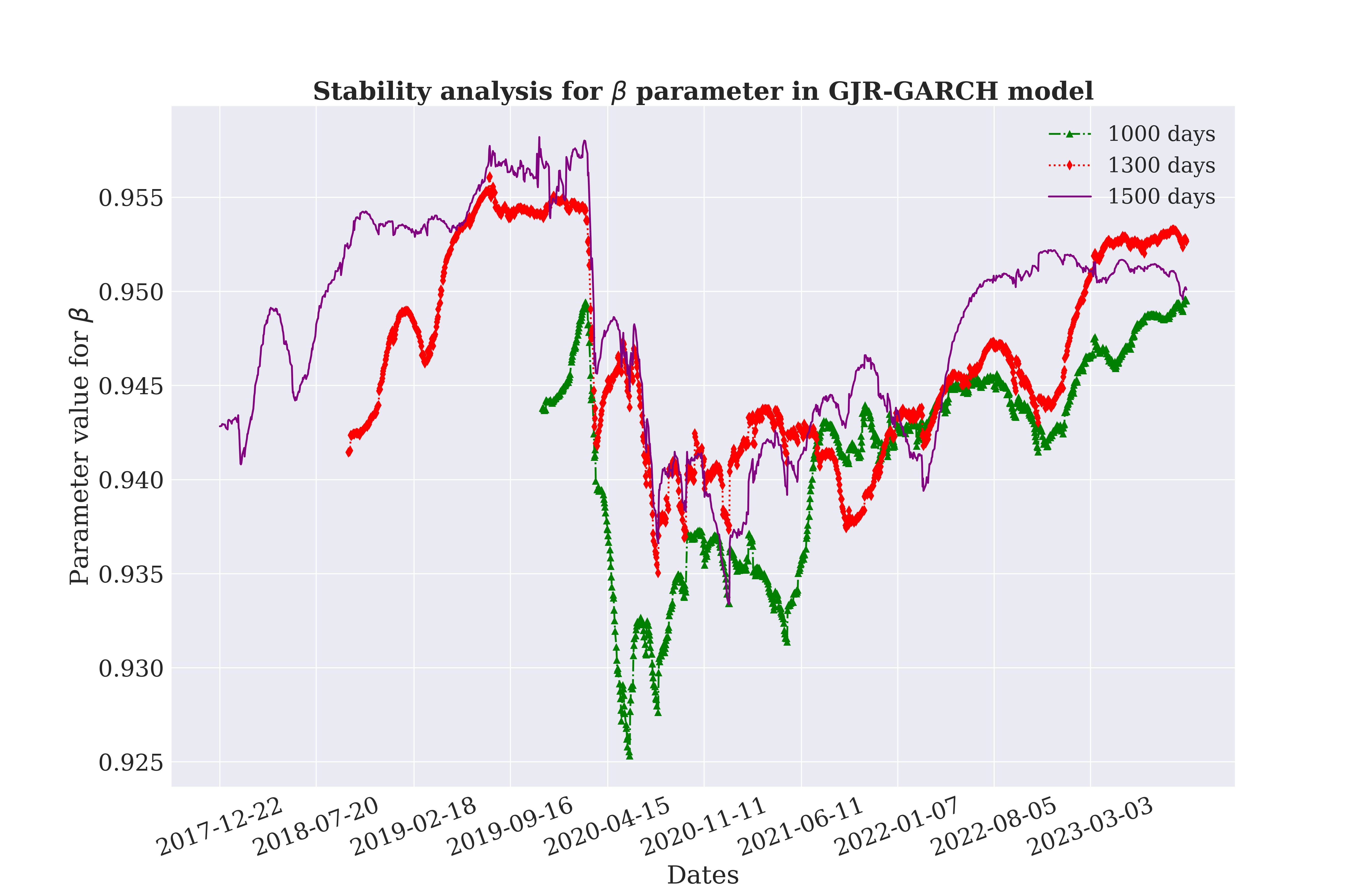}
    \caption{GJR-GARCH $\beta$ parameter stability testing for long testing periods.}
    \label{fig::Paramater_stability_BETA_GJR_1}
\end{figure}

\begin{figure}[H]
    \centering
    \includegraphics[width=1.0\linewidth]{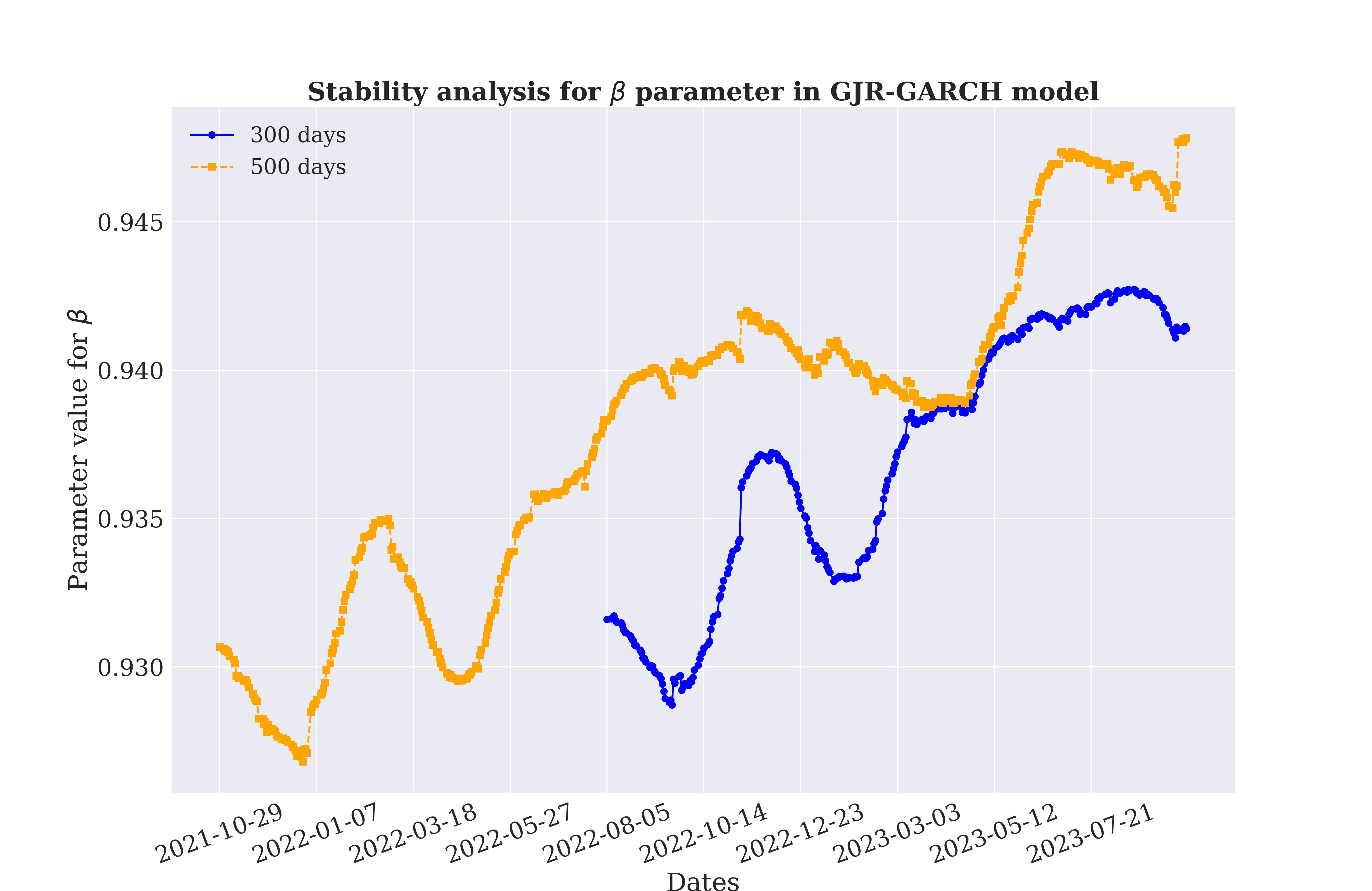}
    \caption{GJR-GARCH $\beta$ parameter stability testing for short testing periods.}
    \label{fig::Paramater_stability_BETA_GJR_2}
\end{figure}

$\beta$ parameter on the other hand shows a small drop in value at the beginning of the COVID-19 pandemic, which supports lower volatility persistence in turbulent times (see Figure \ref{fig::Paramater_stability_BETA_GJR_1}). But it comes back to the higher values after some time (see Figure \ref{fig::Paramater_stability_BETA_GJR_2}).

\begin{figure}[H]
    \centering
    \includegraphics[width=1.0\linewidth]{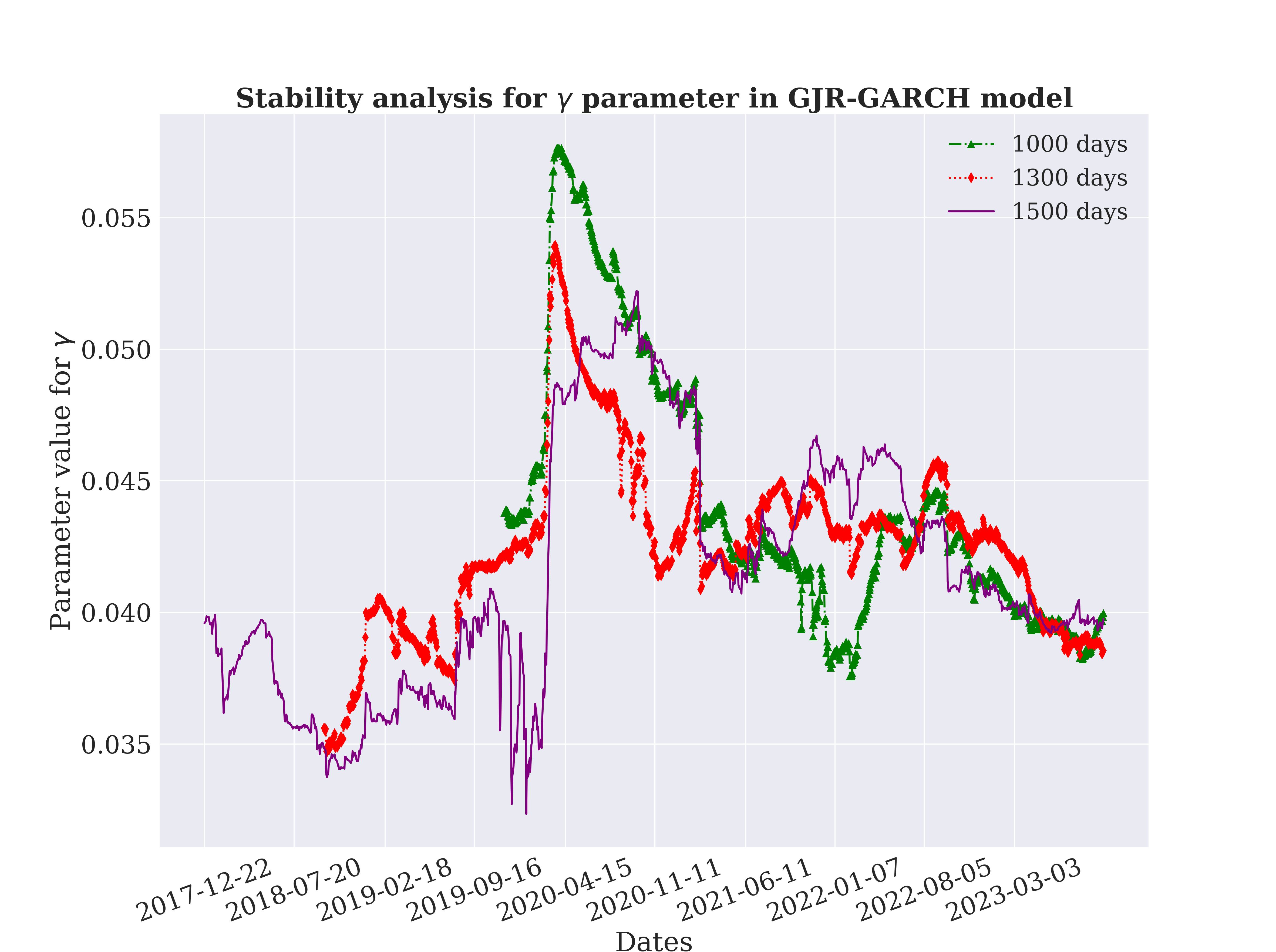}
    \caption{GJR-GARCH $\gamma$ parameter stability testing for long testing periods.}
    \label{fig::Paramater_stability_GAMMA_GJR_1}
\end{figure}

\begin{figure}[H]
    \centering
    \includegraphics[width=1.0\linewidth]{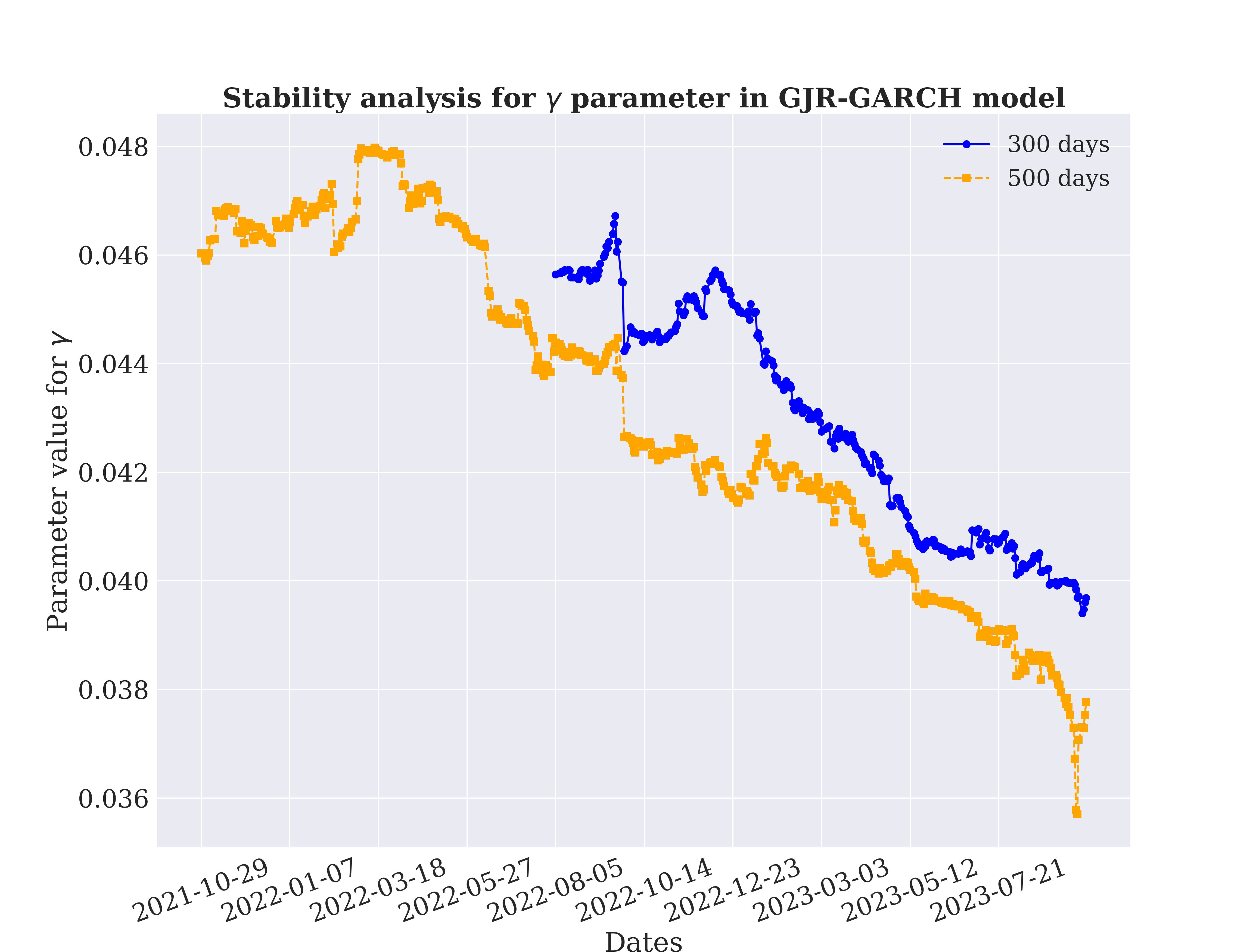}
    \caption{GJR-GARCH $\gamma$ parameter stability testing for short testing periods.}
    \label{fig::Paramater_stability_GAMMA_GJR_2}
\end{figure}

As far as $\gamma$ parameter is concerned we can see a similar situation as in the case of $\alpha$ parameter. Leverage is gaining momentum when the overall level of short-term volatility is rising, which is mostly apparent at the beginning of COVID-19 (see Figure \ref{fig::Paramater_stability_GAMMA_GJR_1}).
 However, this effect fades away as the markets become more stable (see Figure  \ref{fig::Paramater_stability_GAMMA_GJR_2}).

\begin{figure}[H]
    \centering
    \includegraphics[width=1.0\linewidth]{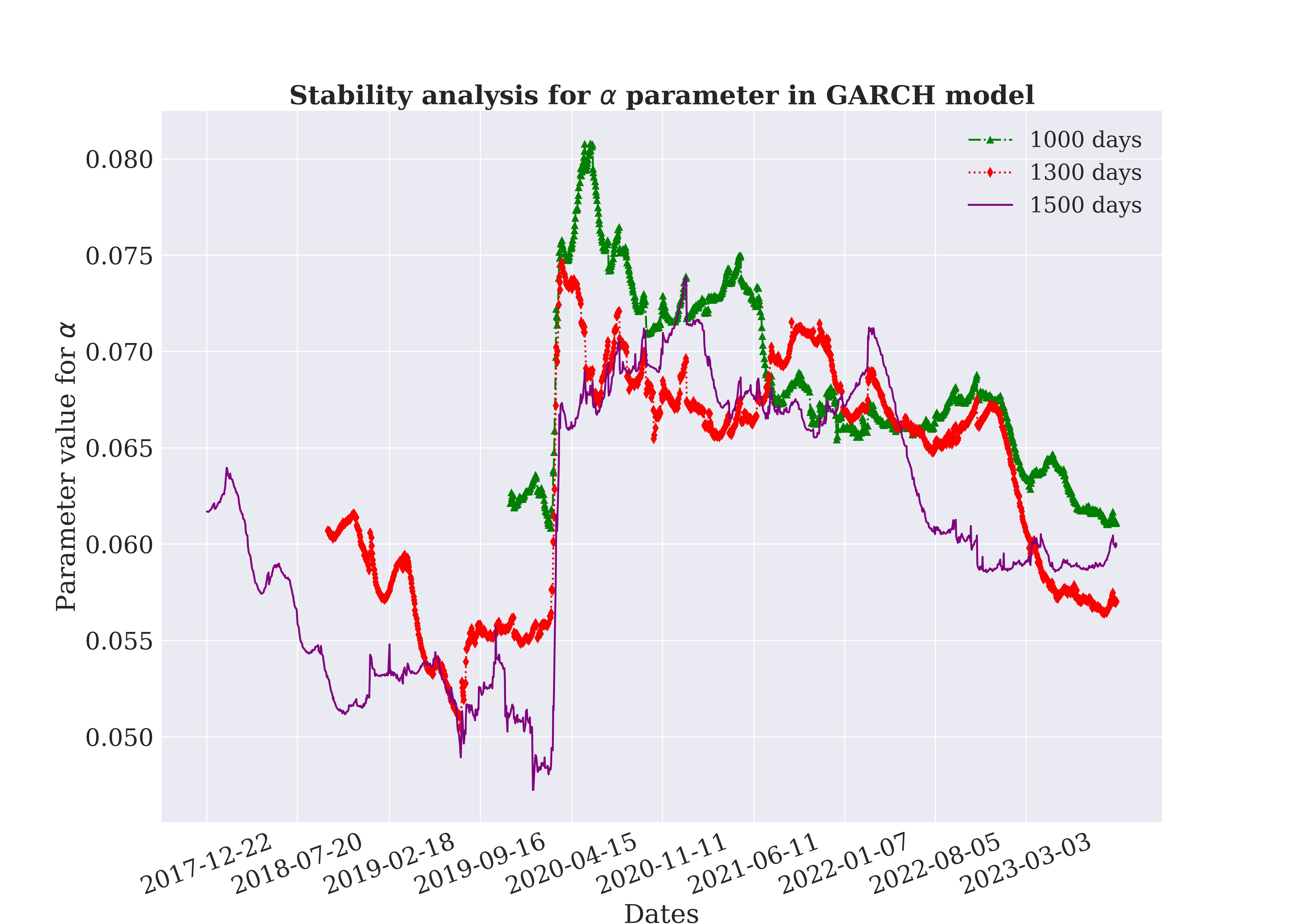}
    \caption{GARCH $\alpha$ parameter stability testing for long testing periods.}
    \label{fig::Paramater_stability_ALPHA_GARCH_1}
\end{figure}

As far as the regular GARCH model is concerned we can see similar behavior of parameters. 
Figure \ref{fig::Paramater_stability_ALPHA_GARCH_1} shows a sharp increase in $\alpha$ parameters across all the testing periods at the beginning of the COVID-19 pandemic. This behavior is expected as $\alpha$ parameter reflects the most recent market occurrences and captures current nervousness in the markets. 

\begin{figure}[H]
    \centering
    \includegraphics[width=1.0\linewidth]{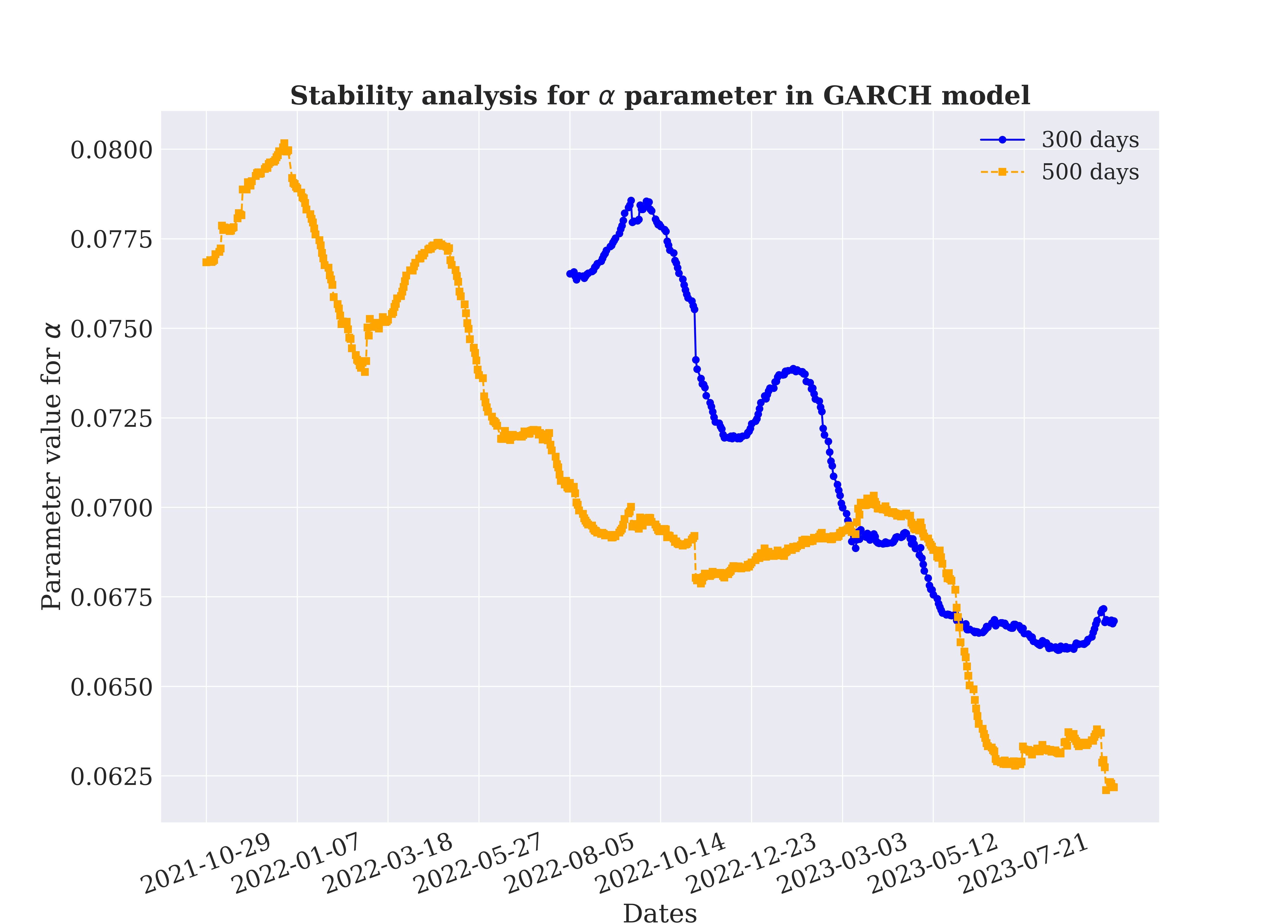}
    \caption{GARCH $\alpha$ parameter stability testing for short testing periods.}
    \label{fig::Paramater_stability_ALPHA_GARCH_2}
\end{figure}

 Next we can see that this effect fades away as markets are getting more stable, see Figure \ref{fig::Paramater_stability_ALPHA_GARCH_2}

\begin{figure}[H]
    \centering
    \includegraphics[width=1.0\linewidth]{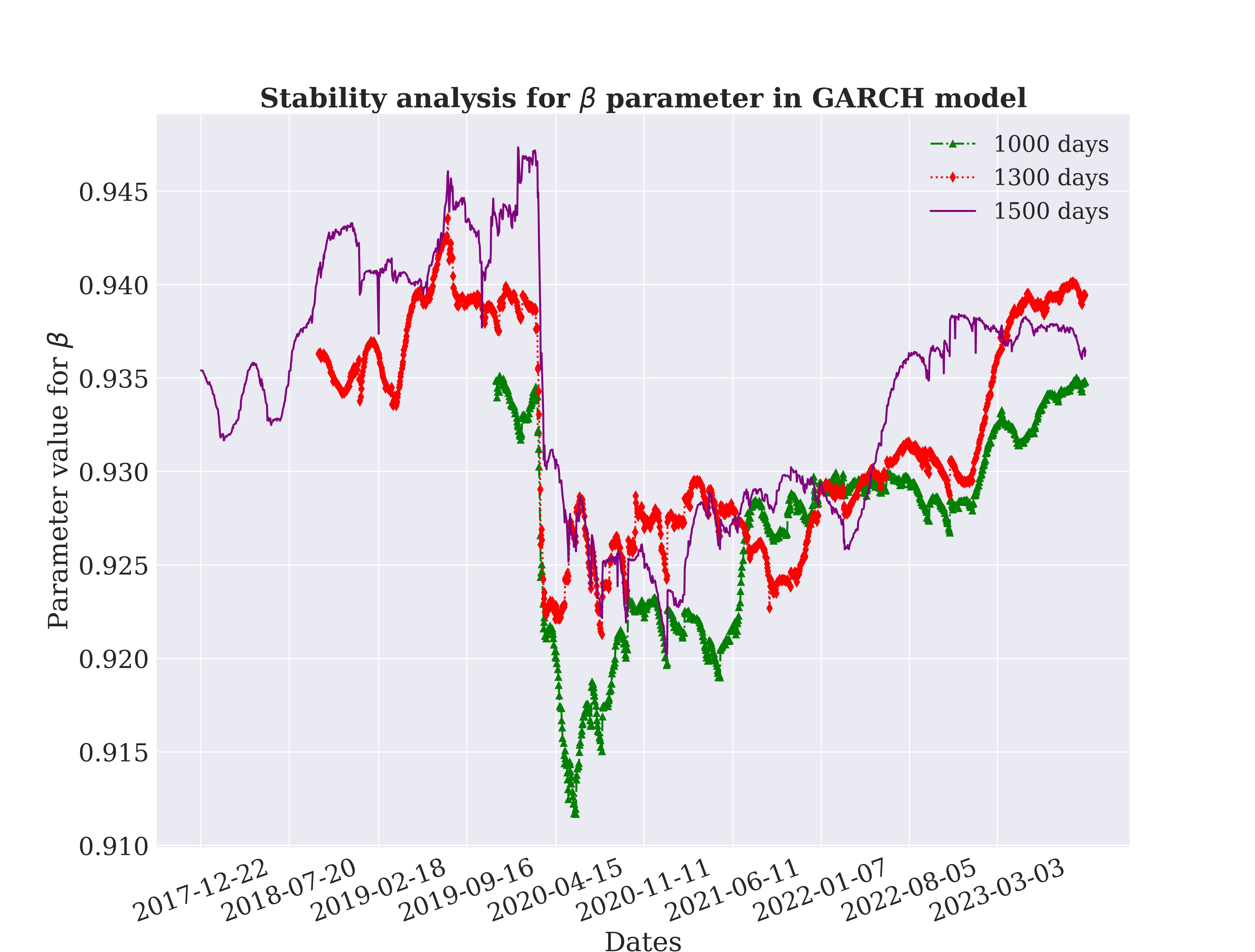}
    \caption{GARCH $\beta$ parameter stability testing for short testing periods.}
    \label{fig::Paramater_stability_BETA_GARCH_1}
\end{figure}

\begin{figure}[H]
    \centering
    \includegraphics[width=1.0\linewidth]{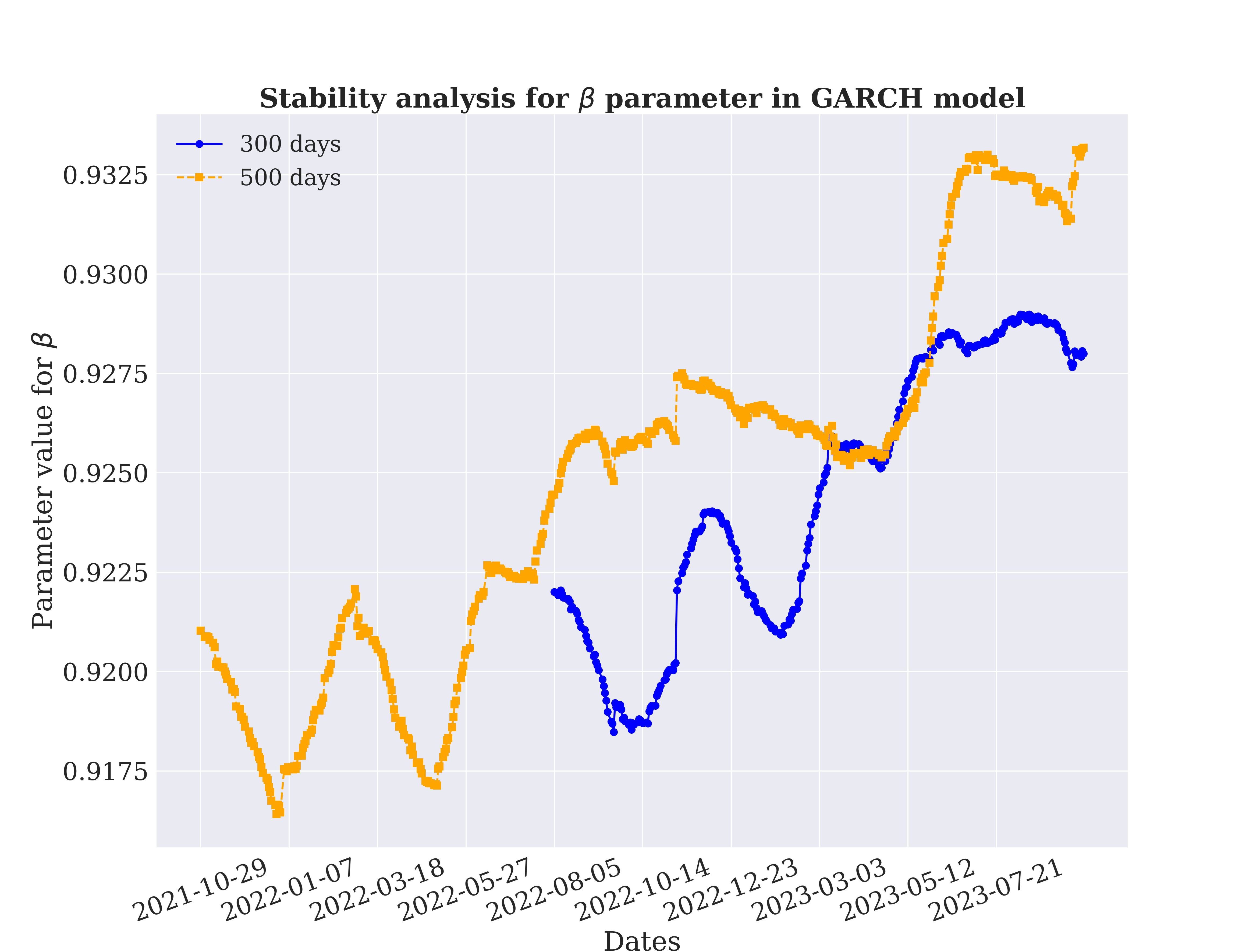}
    \caption{GARCH $\beta$ parameter stability testing for short testing periods.}
    \label{fig::Paramater_stability_BETA_GARCH_2}
\end{figure}

Thus, $\alpha$ parameter behaves in a similar fashion in case of GARCH and GJR-GARCH models. 
Taking a closer at $\beta$ parameter we can also draw similar conclusions as in the case of GJR-GARCH. 
Figure \ref{fig::Paramater_stability_BETA_GARCH_1} shows that the strength of volatility persistence weakens in more turbulent times. And rebounds when the market sentiment is more stable as seen in Figure \ref{fig::Paramater_stability_BETA_GARCH_2}.

{

The rolling-window analysis provides the motivation for introducing the 
MSGARCH specification. In standard GARCH models, the time 
variation in the long-run variance is absorbed through changes in $\omega$, which 
leads to apparent parameter instability. In MSGARCH, the regime-specific 
parameters $(\omega_k, \alpha_k, \beta_k)$ remain constant, while structural changes are 
represented by the latent state process $S_t$ and its filtered probabilities 
$P(S_t = k \mid \mathcal{F}_t)$. 

Consequently, MSGARCH is not expected to materially outperform single-regime 
GARCH models in one-step-ahead coverage—an effect also confirmed in our 
backtesting results. Its advantage is structural rather than short-term: the model 
provides stable regime parameters and shifts all time variation into the transition and 
occupation probabilities, offering a coherent explanation for the parameter 
instability observed in single-regime models.

To further compare the three models in scope we have calculated the maximised log-likelihood together with the
Akaike (AIC) and Bayesian (BIC) information criteria. Table~\ref{tab:ic_models} presents respective resuts. The MSGARCH(2-regime)
model attains the highest log-likelihood among all specifications, indicating the best
overall in-sample fit. However, once the number of parameters is penalised, the
results diverge: MSGARCH achieves the lowest AIC, but the BIC favours the
GJR-GARCH model due to its substantially smaller parameter set. These outcomes
are consistent with the view that MSGARCH improves the structural representation
of volatility dynamics—particularly in capturing regime-specific persistence and
shock transmission—while the marginal improvement in short-horizon fit remains
modest relative to simpler single-regime models.

\begin{table}[ht!]
\centering
\caption{Log-likelihood and information criteria for competing volatility models.
The table reports the maximised log-likelihood values and corresponding information
criteria for three specifications: standard GARCH(1,1), asymmetric GJR-GARCH(1,1),
and a two-regime MSGARCH(1,1) model.}
\label{tab:ic_models}
\begin{tabular}{|l|c|c|c|c|}
\hline
\textbf{Model} & $\boldsymbol{\log L}$ & $\boldsymbol{k}$ & \textbf{AIC} & \textbf{BIC} \\ \hline
GARCH(1,1)          & -10708 & 4 & $-21407$ & $-21382$ \\ \hline
GJR-GARCH(1,1)      & -10718 & 5 & $-21426$ & $\mathbf{-21394}$ \\ \hline
MSGARCH(2-regime)   & -10722 & 9 & -21427 & $-21369$ \\ \hline
\end{tabular}

\vspace{0.7em}
\noindent\textit{Notes.}
\begin{itemize}
    \item $\log L$ — maximised log-likelihood of the model, i.e., 
    $\log L(\hat{\theta}) = \sum_{t=1}^{T}\log f(r_t \mid \hat{\theta})$.
    Higher values indicate a better in-sample fit.
    \item $k$ — the number of estimated parameters in each model.
    For the MSGARCH(2-regime) model this includes two sets of GARCH parameters
    plus the transition probabilities of the Markov chain and the common mean $\mu$.
    \item AIC — Akaike Information Criterion,
    \[
        \mathrm{AIC} = -2\log L + 2k,
    \]
    which penalises model complexity moderately.
    Lower values indicate better model quality.
    \item BIC — Bayesian Information Criterion,
    \[
        \mathrm{BIC} = -2\log L + k \ln T,
    \]
    which penalises complexity more strongly than AIC.
\end{itemize}

\end{table}
\FloatBarrier

Figure~\ref{fig:ms_filt_prob} displays the filtered probability of the 
high-volatility regime $P(S_t = 2 \mid \mathcal{F}_t)$ obtained from the 
two-regime MSGARCH(1,1) model. The results show a clear and intuitive pattern. 
The probability of the high-volatility regime is close to one during the early part 
of the sample, corresponding to the 2007--2009 financial crisis and its aftermath. 
It then collapses to values near zero for a prolonged tranquil period 
characterised by low rates, large-scale monetary accommodation and compressed 
volatility. A second pronounced cluster of high-volatility probabilities emerges 
around March~2020, in line with the market dislocations associated with the 
COVID-19 shock. 

Beyond these two major episodes, regime~2 only appears in short-lived spikes, 
typically associated with inflation-driven repricing and the LIBOR--OIS--SOFR 
transition. These do not trigger a persistent regime change, confirming that the 
structural dynamics of the interest-rate volatility process are well captured by a 
two-regime specification with stable parameters and slowly moving regime 
probabilities.

\begin{figure}[!ht]
    \centering
    \includegraphics[width=1.0\linewidth]{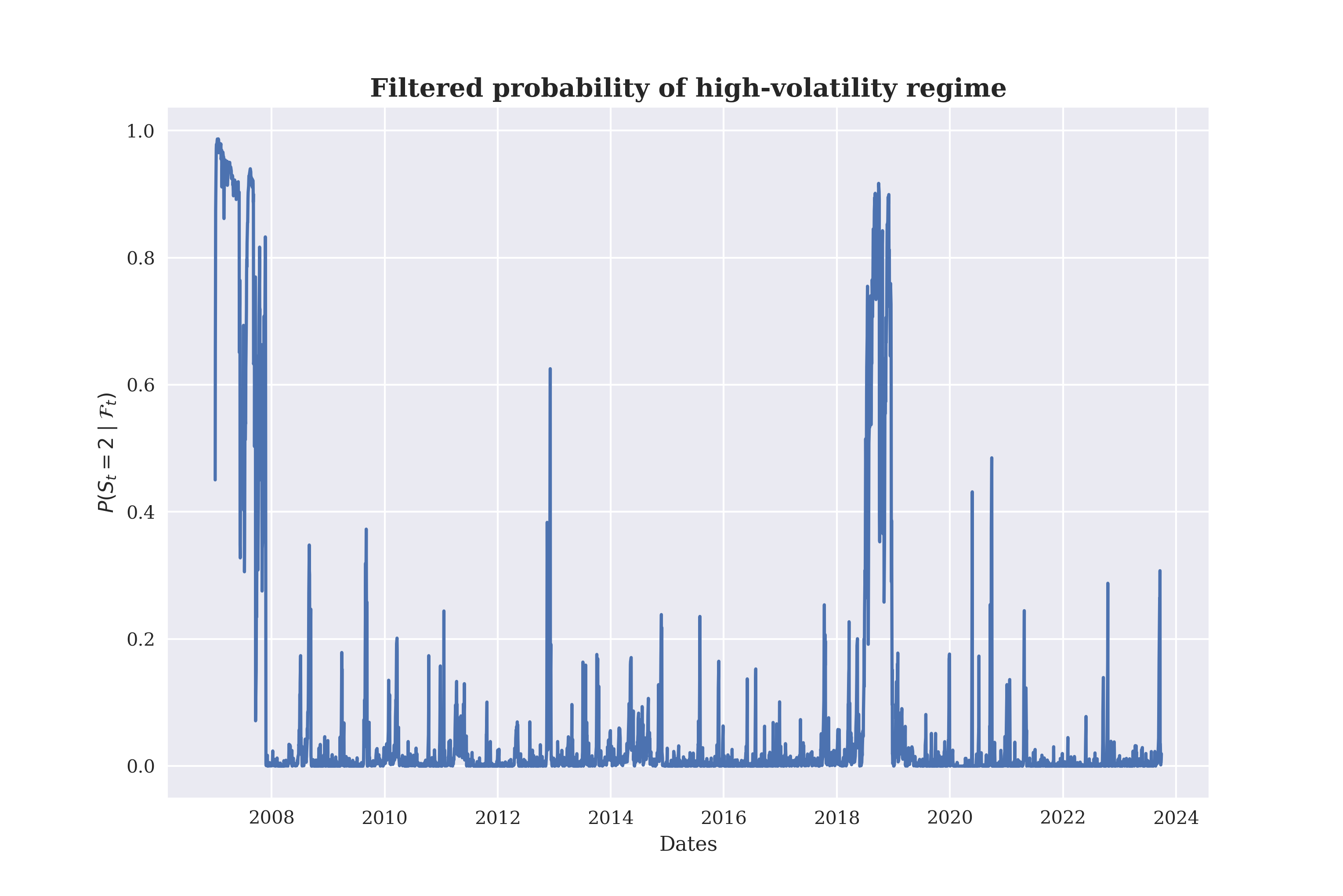}
    \caption{Filtered probability of high volatility regime in MSGARCH model}
    \label{fig:ms_filt_prob}
\end{figure}
\FloatBarrier

In the rolling-window backtest in Section \ref{sec:backtest} we re-estimated the
GARCH(1,1) and GJR-GARCH(1,1) models on moving samples and tracked the
evolution of their parameters. The results show that the short-run dynamics 
parameters ($\alpha$, $\beta$, and $\gamma$ in the GJR-GARCH specification) remain
remarkably stable over time, exhibiting only minor fluctuations. In contrast, the
intercept parameter $\omega$, which determines the unconditional variance
$\omega / (1 - \alpha - \beta)$, varies substantially across windows. As a consequence,
the implied long-run variance shows pronounced instability. This pattern is fully
consistent with the Nyblom stability test, which rejects joint parameter stability for
both single-regime models: even though $\alpha$, $\beta$ and $\gamma$ behave
consistently, the strong variation in $\omega$ renders the overall specification
non-stationary over the full 2007--2023 period.

The MSGARCH model provides a structural resolution of this issue. Once
estimated on the full sample, its regime-specific parameters
$(\omega_k, \alpha_k, \beta_k)$ remain stable, and the time variation previously picked
up by $\omega$ in the single-regime GARCH is now absorbed by the filtered regime
probabilities $P(S_t = k \mid \mathcal{F}_t)$. This decomposition confirms that the
apparent parameter instability of standard GARCH models is in fact the result of
latent regime changes in the volatility process rather than true structural drift in the
GARCH coefficients themselves.

For the MSGARCH specification, the notion of ``parameter stability'' is different.
Hence, re-estimating a full MSGARCH model on each rolling window would therefore be both numerically demanding and conceptually
misleading: the purpose of MSGARCH is precisely to separate stable regime
parameters from time-varying regime occupancy.

Instead, we estimate MSGARCH once on the full sample and apply a Hamilton-style
filter to obtain the filtered regime probabilities $P(S_t = k \mid \mathcal{F}_t)$ for
each day in the sample. The Hamilton-style filter is the standard recursive Bayesian algorithm used to infer the latent
Markov state in regime-switching models. At each time step, it updates the probability of being
in each regime by combining the one-step-ahead transition prediction with the likelihood implied
by the current observation. This procedure yields filtered probabilities 
$P(S_t = i \mid \mathcal{F}_t)$ that capture the real-time regime classification generated by the
estimated MSGARCH dynamics.

The resulting time series of the high-volatility regime
probability $P(S_t = 2 \mid \mathcal{F}_t)$ aligns closely with major market events:
it spikes during the 2008--2009 financial crisis and the March 2020 COVID shock,
and remains persistently low during the mid-2010s low-volatility period. Around the
transition from the LIBOR to the OIS and SOFR regimes, we observe gradual shifts
in the dominant state rather than abrupt parameter changes, which is exactly the
behaviour MSGARCH is designed to represent.

From the backtesting perspective, this implies that MSGARCH does not aim to
outperform single-regime GARCH models in one-step-ahead coverage---indeed, all
models achieve very similar empirical coverage rates. Its main advantage lies in the
structural decomposition of volatility into persistent low- and high-volatility regimes
with stable parameters, while the observed non-stationarities are absorbed by the
time-varying regime probabilities rather than by drifting GARCH coefficients. This
provides a coherent explanation for the parameter instability detected in the
single-regime models and links it directly to observable macro-financial episodes.
}

{

\section{Impact of Collateralisation Regimes on GARCH Dynamics}
\label{sec:regime_effects}

The full sample analysed in this study (2007--2023) spans three distinct
collateralisation regimes in USD interest rate markets: the pre-crisis
LIBOR-based framework, the post-crisis OIS discounting regime, and the most recent
transition to the SOFR benchmark. Each of these regimes is characterised by
different liquidity conditions, funding conventions, and term-structure dynamics,
all of which may affect the behaviour of conditional volatility models. In this
section we examine how GARCH-type parameters evolve across these structural
breaks and how regime changes influence volatility persistence and shock
sensitivity. As shown in Sections \ref{sec:msgarch} and \ref{sec:backtest}, these shifts are also a key driver of the
instability of the intercept parameter $\omega$ in single-regime GARCH models, and
they provide economic motivation for introducing a Markov--switching GARCH
(MSGARCH) specification capable of structurally capturing such market transitions.

\subsection*{LIBOR regime (2007--2010)}
The pre-crisis period is characterised by elevated uncertainty and pronounced
volatility clustering, particularly around the 2007--2008 financial crisis. GARCH
estimations for this subsample exhibit the highest persistence, with $\beta$ close
to unity, and relatively smaller contributions from recent shocks ($\alpha$). This
pattern indicates that, in an environment dominated by funding stress and
structural dislocations in the interbank market, volatility adjustments occur
gradually and are dominated by long-memory effects. The leverage term in the
GJR-GARCH model also exhibits higher significance during this period, reflecting
the strong asymmetry in interest-rate movements observed during the crisis.

\subsection*{OIS discounting regime (2010--2019)}
Following the industry-wide transition to OIS discounting, driven by mandatory
central clearing and changes in collateralisation practices, the behaviour of swap
rates becomes more stable and more tightly linked to overnight rates rather than
unsecured term funding. In this regime, GARCH parameters indicate a noticeable
shift: volatility persistence ($\beta$) declines, while the ARCH coefficient ($\alpha$)
increases, signalling a stronger sensitivity to short-term innovations. This reflects a
market in which new information is incorporated more rapidly into swap rate
dynamics. The leverage effect, while still present, becomes weaker relative to the
LIBOR era, consistent with reduced credit-risk asymmetry in the underlying rates.

\subsection*{SOFR regime (2019--2023)}
The most recent transition to SOFR introduces another structural break, reflecting
the replacement of a term unsecured benchmark with a nearly risk-free rate. In this
period, both GARCH and GJR-GARCH models display lower unconditional volatility and
moderate persistence. Parameter stability improves, and the leverage coefficient in
the GJR-GARCH specification loses statistical significance in several windows,
supporting the conclusion that asymmetry plays a smaller role when the driving
benchmark is a secured overnight rate. However, during episodes of market stress
(e.g.\ the COVID-19 shock in early 2020), we observe temporary surges in $\alpha$
and drops in $\beta$, reflecting the rapid rise in short-term volatility typical of
crisis dynamics.

\medskip
\noindent\textbf{Integration with the MSGARCH perspective.}
\smallskip

Across the three regimes, a clear pattern emerges: (i) volatility persistence is highest
in the LIBOR era, declines in the OIS period, and stabilises at moderate levels in
the SOFR regime; (ii) short-term shock sensitivity increases progressively, with
$\alpha$ becoming more responsive in the post-crisis environment; (iii) leverage
effects weaken over time, reinforcing the view that asymmetry in interest-rate
dynamics has diminished as credit-sensitive benchmarks were replaced with
near risk-free alternatives. From the perspective of single-regime GARCH models,
these shifts manifest as strong time variation in the intercept parameter $\omega$
and in the implied long-run variance, a point confirmed by both the rolling-window
exercise and the Nyblom stability test.

The MSGARCH specification provides a structural interpretation of these
collateral-driven regime shifts. Instead of allowing $\omega$ to drift across windows,
the model maintains stable regime-specific parameters
$(\omega_k,\alpha_k,\beta_k)$ and captures transitions between the LIBOR, OIS and
SOFR eras through changes in the filtered regime probabilities
$P(S_t=k \mid \mathcal{F}_t)$. The high-volatility regime dominates during the
financial crisis and the COVID-19 shock, while the low-volatility regime remains
persistent throughout the mid-2010s OIS period. Around the SOFR transition the
model detects short-lived increases in the probability of the high-volatility state,
reflecting the repricing uncertainty associated with the benchmark change. This
evidence supports the conclusion that collateralisation conventions are a
quantitatively important driver of volatility regime switches and that MSGARCH 
offers a coherent and economically grounded representation of these structural
dynamics.

}

\section{Considerations on Total Market Variance}
\label{sec:TotalVar}

%Our analysis so far was based on relating GARCH filter to the at-the-money swaption volatility only. 
Thus far, our analysis has focused solely on the relationship between the GARCH filter and at-the-money swaption volatility. However, this approach does not capture the whole information that the market conveys about the volatility of the underlying instrument. From the point of view of volatility (or option trading), we should also consider a skew. A good candidate for this purpose could be the Interest Rate Swap Volatility Index that used to be quoted on the Chicago Board of Options Exchange, see \cite{SrVix}. The SRVIX calculation resembles VIX computations and is based on a simplified option-implied pricing formula for a hypothetical variance swap contract on forward swap rates:

\begin{align}
  SRVIX_t(M,T) = 100^2 \left(\frac{2}{PVBP_t\times M} \sum_{\substack{i:K_i<R_t}} {SW^R_t\left(K_i,M,T\right)\Delta K_i} + \right. \nonumber \\ 
\left. \sum_{\substack{i:K_i\geq R_t}} {SW^P_t}(K_i,M,T) \Delta K_i \right)^{0.5},
\end{align}
\noindent\textrm{where:}
$t$ - time of calculation,
$T$ - swap tenor,
$R_{t}$ - at-the-money forward swap rate,
$PVBP_{t}$ - price value of a basis point (swap annuity),
$SW_{t}^{R}$ - price of corresponding receiver swaption,
$SW_{t}^{P}$ - price of corresponding payer swaption,
$K_{i}$ - highest Out-of-The-Money strike available,
$\Delta K_{i}$ - strike increment
\newline

Index formula covers only 11 swaptions with respective ITM and OTM strikes, see details in \cite{SrVix}.
What we postulate here is that the annualized square root of GARCH filter can mimic the history of SRVIX. In order to check upon this claim we perform two analyses:
\begin{enumerate}
    \item We have price 1000 European Swaptions, i.e., 500 payers and 500 receivers with volatilities implied by the SABR model (Stochastic Alpha Beta Rho), see \cite{Hagan}, that has been calibrated to the available swaption volatility cube. SRVIX index realizations have been obtained according to the formula presented above.
    Model free replication of fair variance swap rate assumes usage of a continuum of swaptions, see \cite{Hilpisch}. Hence for this purpose we discretionarily use 1000.

    \item We use strikes predefined in the CBOE specifications and read volatilities directly off the surface and similarly apply relevant index formula.
\end{enumerate}

\begin{figure}[H]
    \centering
    \includegraphics[width=1.0\linewidth]{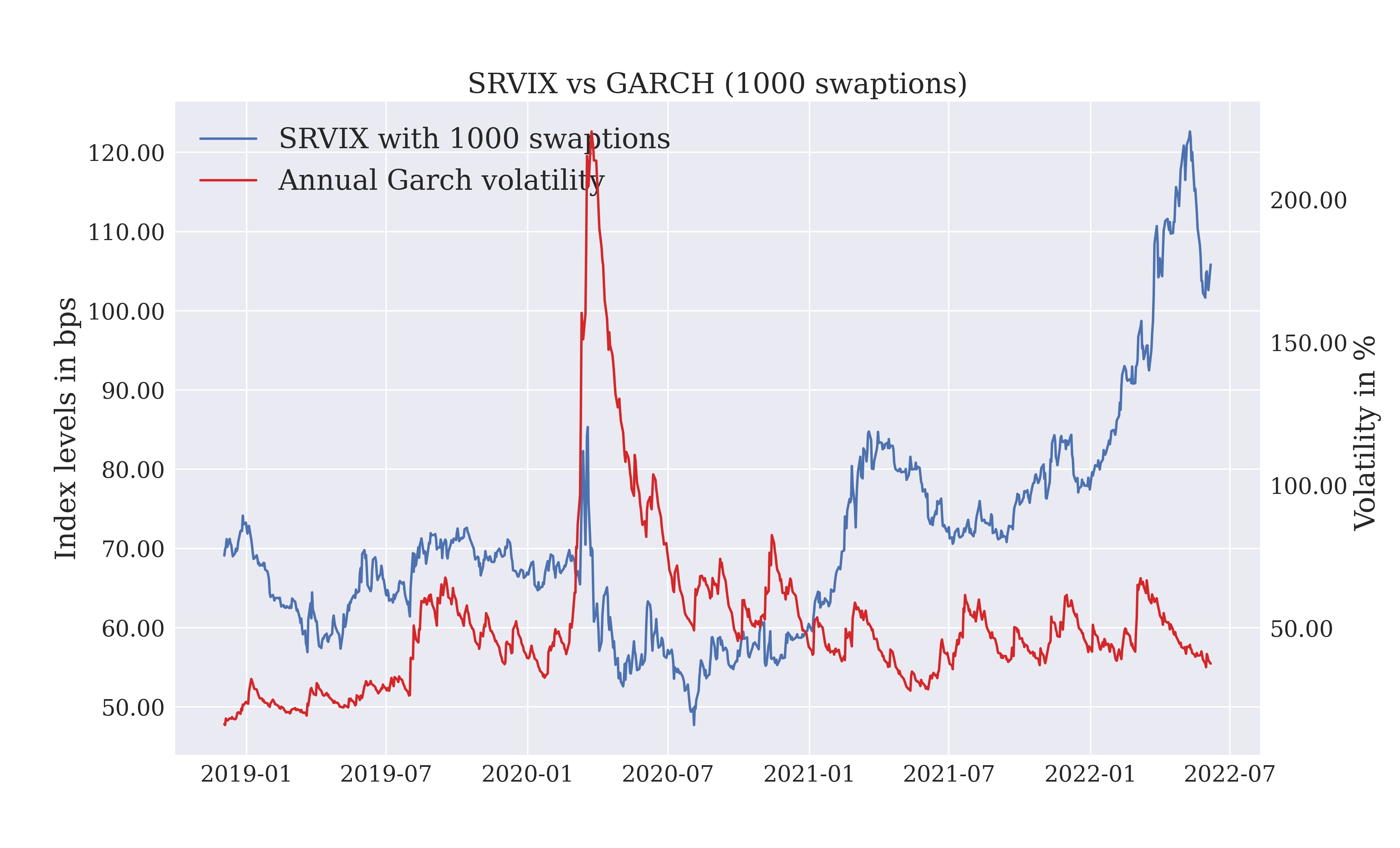}
    \caption{SRVIX versus GARCH filter - 1000 swaptions. Annual GARCH volatilities and SRVIX replication diverge.}
    \label{fig::SRVIX_vs_GARCH_filter_1000swaptions}
\end{figure}

Figure \ref{fig::SRVIX_vs_GARCH_filter_1000swaptions} presents relevant results for the first case analyzed, which shows that SRVIX calculated with 1000 swaptions does not trend well with GARCH volatility.

\begin{figure}[H]
    \centering
    \includegraphics[width=1.0\linewidth]{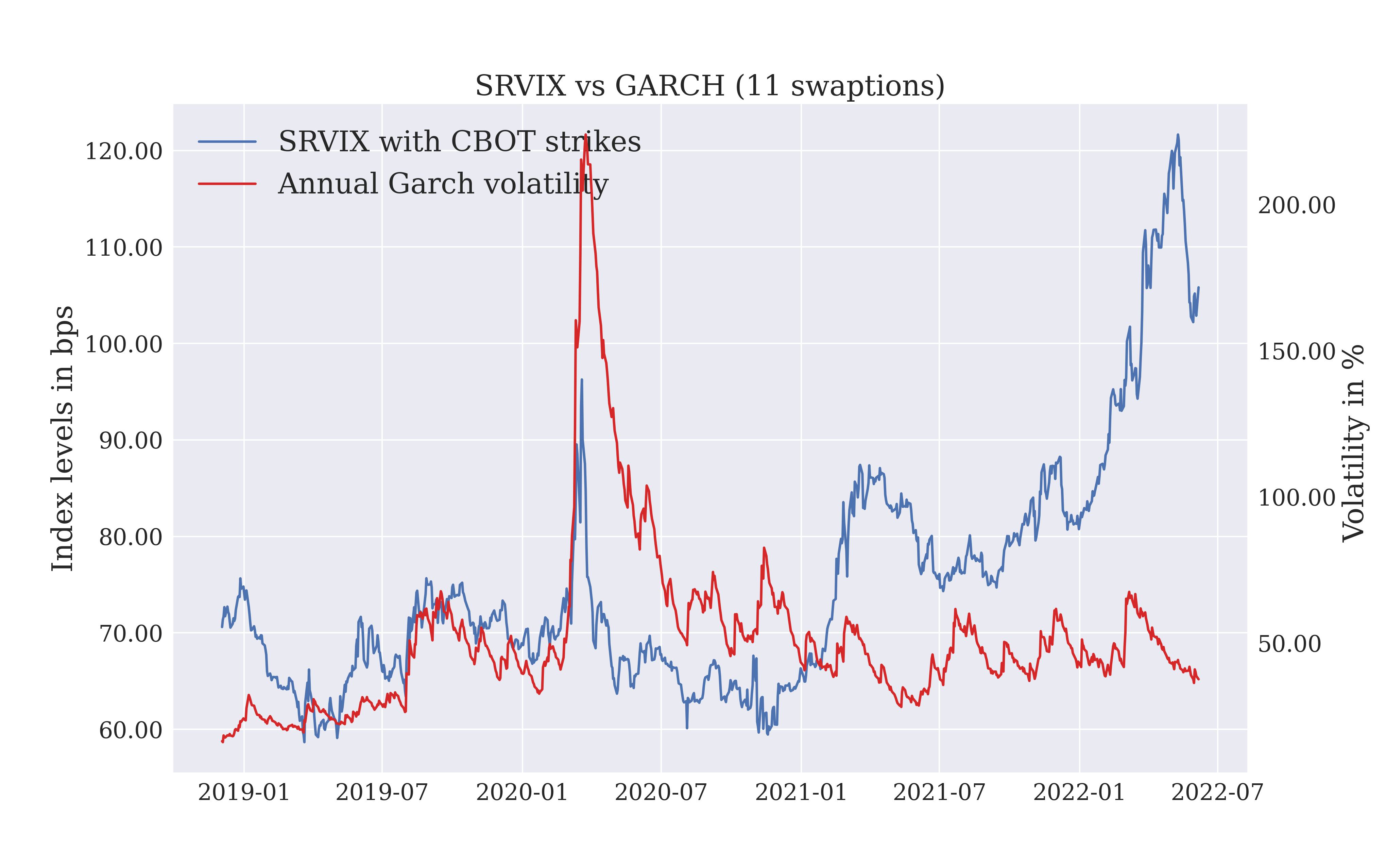}
    \caption{SRVIX versus GARCH filter - 11 swaptions. Similarly as in exmaple above annual GARCH volatilities and SRVIX replication diverge.}
    \label{fig::SRVIX_vs_GARCH_filter_11Swaptions}
\end{figure}

GARCH filter does not follow SRVIX index calculated according to the white-paper formula and the implied correlation for the whole period equals to $-14\%$.
Next, we have calculated index realizations exactly as specified by CBOE in \cite{SrVix}, i.e, with 11 swaptions and volatilities read off the surface. 
Figure \ref{fig::SRVIX_vs_GARCH_filter_11Swaptions} presents relevant results for this case.

Similarly as in the first case, we can see that GARCH filter and SRVIX do not trend together. Implied correlation for the whole tested period equals to 1.5\%.
Finally, we turned to the article by Carr and Madan (see \cite{CarrMadan}) and J.P. Morgan Derivatives Research (see \cite{JPMorgan}) where the authors provide comprehensive treatment of a Variance Swap contract. The point of the analysis is to set up a portfolio of European-styled options that has constant exposure to the volatility, understood as a constant difference between implied and realized variance of an underlying. It turns out that such a portfolio must be weighted by the inverse strike squared. In the next steps, the authors derive static log-contract replication via a continuum of European call and put options and finally come up with a forward cost of such a portfolio expressed as
\begin{align}
K_{\text{VAR}}^2 = \frac{2e^{rT}}{T}\left[\int_0^{F_0}\frac{P_0(K)}{K^2} \, dK + \int_{F_0}^{\infty}\frac{C_0(K)}{K^2} \, dK \right].
\end{align}

Hence, with little adjustment for annuity, we have applied this formula to replicate SRVIX index. 

\begin{figure}[H]
    \centering
    \includegraphics[width=1.0\linewidth]{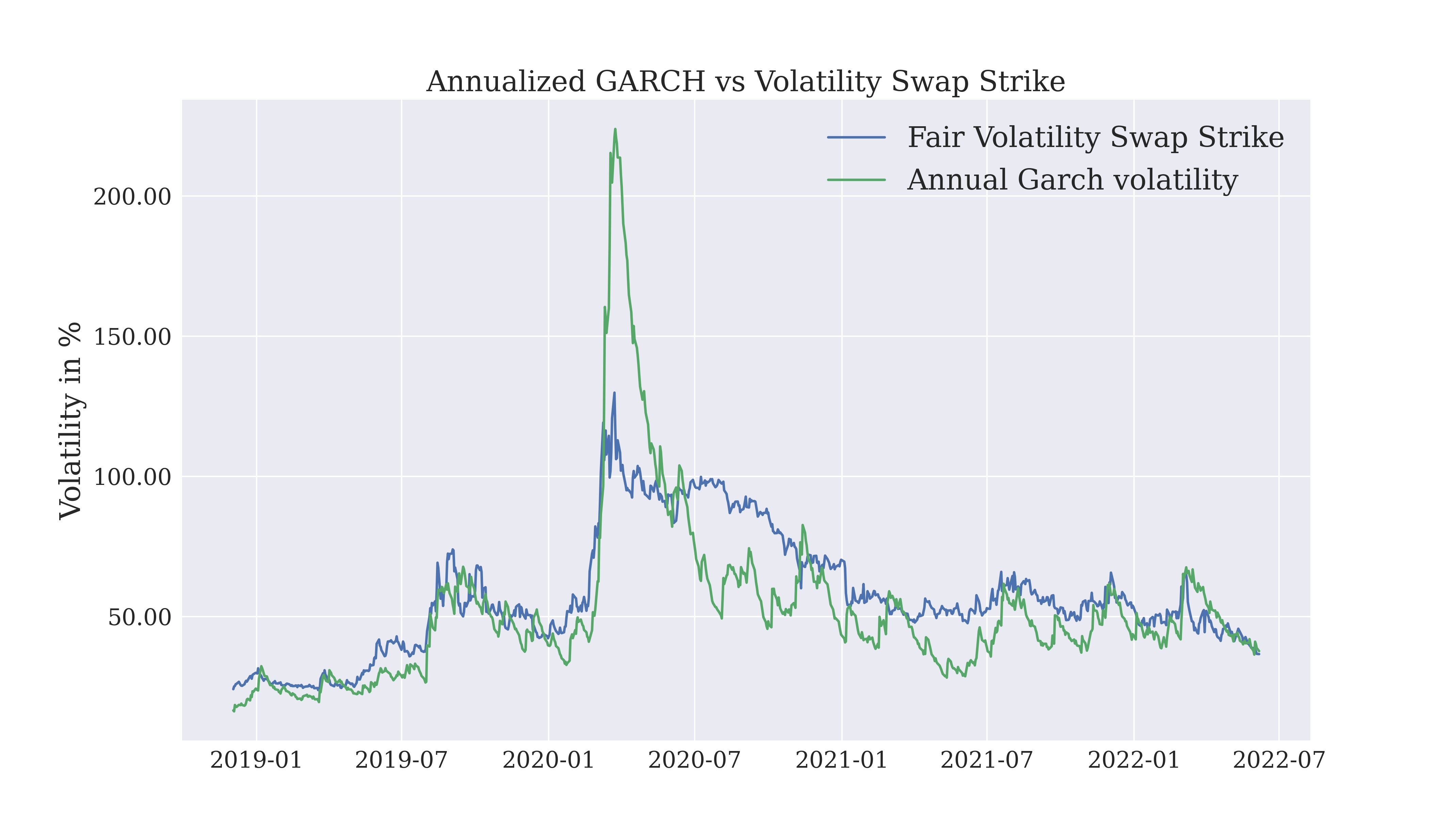}
    \caption{Annualized GARCH filter versus Fair Volatility Strike. This time replicated SRVIX and annual GARCH volatilities trend together.}
    \label{fig::Annualized_GARCH_vs_VolStrike}
\end{figure}

Figure \ref{fig::Annualized_GARCH_vs_VolStrike} shows that SRVIX index calculated according to the Fair Volatility Strike formula trends closely with GARCH filter.
Thus, GARCH filter is able to capture total market volatility and not only ATM.
{
Finally we note that, the MSGARCH filter may also be compared directly with the SRVIX index over the
SOFR era. Figure~\ref{fig:MS_SRIVX_2019_2022} shows respective results. 
A closer comparison of the MSGARCH filter with the SRVIX index over the period
2019-2022 reveals an important nuance. For most of this window—particularly the
post-2020 SOFR transition period—the mixed MSGARCH variance tracks SRVIX very
closely and behaves similarly to the standard GARCH and GJR-GARCH filters. This is
expected: after the initial transition to SOFR the interest-rate market exhibits relatively
stable volatility dynamics, and the filtered probability of the high-volatility regime
remains close to zero for extended periods.

However, during the sharp volatility spike associated with the COVID-19 shock
(March-April 2020), the MSGARCH filter briefly diverges from the single-regime GARCH
estimates. The model detects a short-lived jump in the probability of the 
high-volatility regime and increases the mixed variance accordingly. This behaviour is
consistent with the forward-looking nature of option-implied volatility and with the
abrupt rise in uncertainty captured by the SRVIX index.

Outside that episode, regime-switching is rarely activated. As a result, the MSGARCH
filter reverts quickly to its low-volatility regime and becomes nearly indistinguishable
from a standard GARCH process for most of 2020-2022. The similarity between the
models over this period therefore reflects not a limitation of MSGARCH, but rather the
underlying stability of the SOFR-based environment once market functioning normalised
after the COVID shock.

Overall, the period 2019-2022 illustrates the central insight of our analysis:
MSGARCH provides additional structural information during genuine regime shifts
(e.g., the COVID spike), but aligns with single-regime GARCH models during tranquil
phases. This reinforces the interpretation that MSGARCH excels in identifying 
structural breaks rather than in improving short-horizon variance tracking.

\begin{figure}[!ht]
    \centering
    \includegraphics[width=1.0\linewidth]{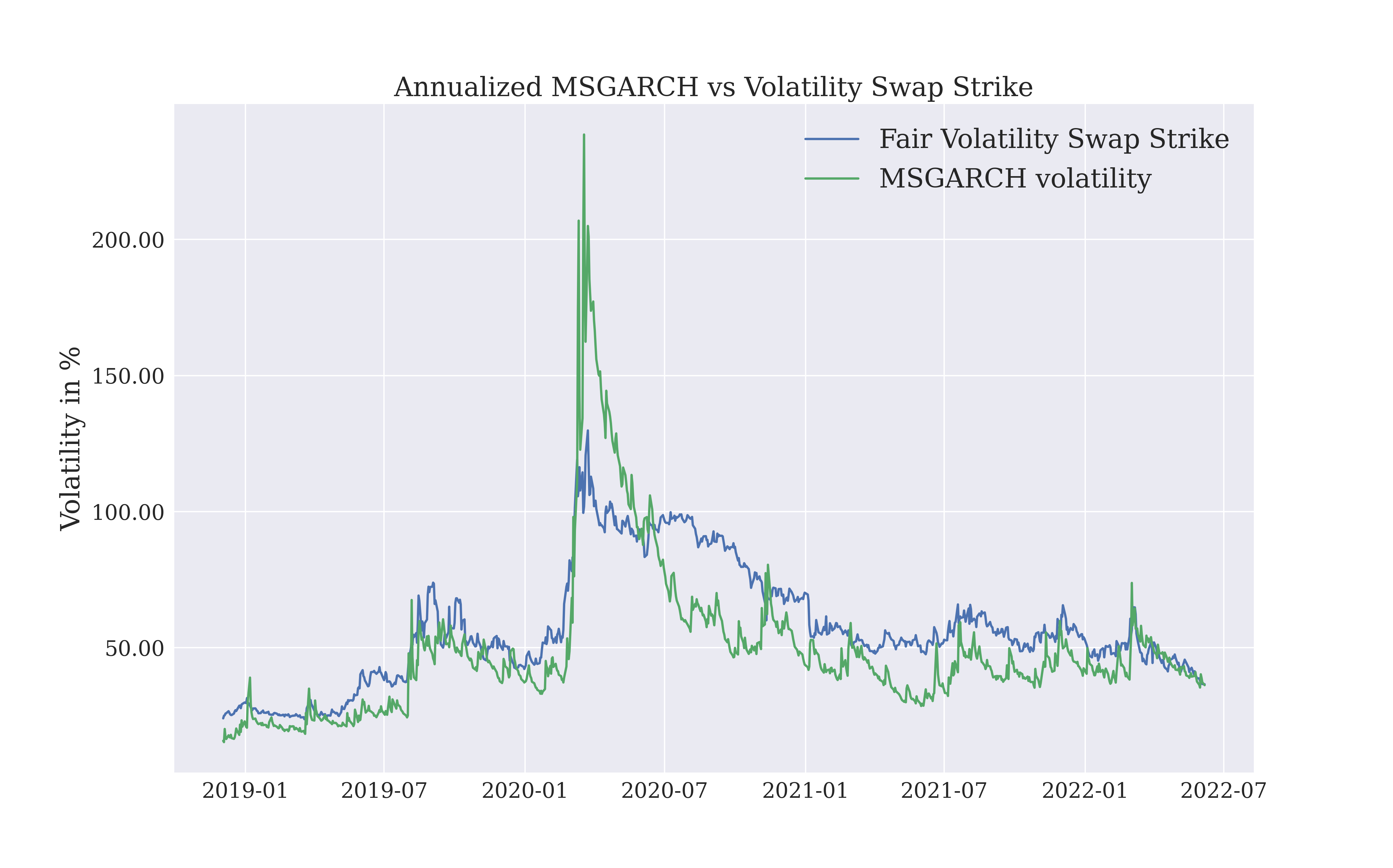}
    \caption{Annualized MSGARCH filter versus Fair Volatility Strike. This time replicated SRVIX and annual MSGARCH volatilities trend together.}
    \label{fig:MS_SRIVX_2019_2022}
\end{figure}
\FloatBarrier

}

{
\section{SRVIX in the literature and contribution of this study}
\label{sec:SRVIX_contribution}

The Swap Rate Volatility Index (SRVIX) was introduced by the Chicago Board Options
Exchange (CBOE) as an interest-rate analogue of the equity VIX index, with the aim of
providing a model–free measure of expected swap rate volatility over a fixed horizon, see
\cite{SrVix} \footnote{We refer to the original CBOE white paper for
implementation details of the index construction.}. The SRVIX methodology relies on
variance swap style replication using a cross-section of payer and receiver swaptions across
a range of out-of-the-money strikes. In spirit, it is closely related to the option-based
approach to fair variance swap rates discussed by Hilpisch \cite{Hilpisch} and the
log-contract replication framework of Carr and Madan \cite{CarrMadan} as well as the
J.P.~Morgan derivatives research note on variance swaps \cite{JPMorgan}.% (cf. \cite[6--8]{}).

Despite its conceptual similarity to VIX, the use of SRVIX in the academic literature
remains relatively limited, especially in the context of validating parametric volatility
models for interest rate swaps. Existing studies typically treat SRVIX as a stand-alone
market-based volatility indicator, or as an input to trading and hedging strategies, rather
than as a benchmark for conditional variance filters. To the best of our knowledge, there is
no systematic analysis linking GARCH-type volatility dynamics of at-the-money (ATM)
forward swap rates to the information content of SRVIX over a long sample spanning
different collateralisation regimes (LIBOR, OIS, SOFR).

The present study contributes to this strand of the literature in two ways. First, we
show that a simple GARCH-type conditional variance specification fitted to daily log
changes of the ATM forward swap rate can produce volatility filters which track the
history of SRVIX reasonably well when the index is computed using a fair volatility strike
formula based on a continuum of payer and receiver swaptions, following the option-based
variance swap replication of \cite{CarrMadan,JPMorgan}. Second, by comparing GARCH
filters not only to ATM implied swaption volatility but also to SRVIX, we effectively
benchmark the model against a measure of \emph{total} market variance rather than a
single point on the volatility surface. This dual validation (ATM implied volatility and
SRVIX) highlights that standard and asymmetric GARCH specifications can be used as
practical, parsimonious tools to approximate both local (ATM) and global (surface-wide)
volatility information embedded in interest rate option markets.
Moreover, we show that incorporating regime dependence through a two-state MSGARCH 
specification provides an additional interpretative layer by linking SRVIX dynamics
 to shifts between low- and high-volatility regimes. The MSGARCH filter remains closely
  aligned with SRVIX during tranquil SOFR-era conditions, while reacting more sharply 
  during structural volatility spikes such as the COVID shock, confirming its value as a
   regime-sensitive extension of standard GARCH models.
}

{
\section{Conclusions}
\label{sec:concl}

This study analysed the conditional volatility of the USD 1Y\,$\times$\,10Y forward
swap rate using a hierarchy of GARCH-type models, beginning with standard
GARCH and GJR-GARCH specifications (Sections \ref{sec:GarchFamily}, \ref{sec:modeltest}, \ref{sec:msgarch}) and extending to
heavy-tailed innovations and a MSGARCH
model (Section \ref{sec:msgarch}). Several key findings emerge from the full 2007--2023 sample.

\medskip
\noindent\textbf{(1) Standard GARCH and GJR-GARCH capture short-run volatility dynamics.}
Both models eliminate autocorrelation and remaining ARCH effects in the residuals,
and the GJR specification appropriately captures the modest leverage effect identified
in Section \ref{sec:hypotest}. Rolling-window backtesting (Section \ref{sec:backtest}) confirms that one-day-ahead
forecast coverage remains close to the nominal 95\% level, demonstrating that these
models perform well for short-horizon volatility prediction.

\medskip
\noindent\textbf{(2) Parameter instability is driven primarily by the long-run variance.}
Nyblom tests (Section \ref{sec:nyblom}) show that the short-run parameters
$(\alpha,\beta,\gamma)$ remain stable, while the intercept parameter $\omega$ exhibits
strong time variation. This pattern is consistent with structural changes in the
fixed-income market and with the regime interpretation in Section \ref{sec:regime_effects}.

\medskip
\noindent\textbf{(3) Heavy-tailed innovations improve statistical fit but do not restore stability.}
Student-$t$, GED and Johnson~$SU$ innovations (Section \ref{subsec:gof_summary}) substantially improve
likelihood and goodness-of-fit relative to the Gaussian case, with Johnson~$SU$
performing best. Nevertheless, unconditional variance remains non-stationary,
indicating that distributional flexibility alone cannot accommodate the structural breaks
present in the sample.

\medskip
\noindent\textbf{(4) MSGARCH provides a coherent structural explanation of volatility shifts.}
The two-regime MSGARCH model (Section \ref{sec:backtest}) achieves the highest log-likelihood and
lowest AIC, with regime-specific parameters $(\omega_k,\alpha_k,\beta_k)$ remaining
stable. Time variation in unconditional volatility is captured by filtered regime
probabilities rather than drifting GARCH parameters. These probabilities align closely
with major market events and benchmark transitions discussed in Section \ref{sec:regime_effects}, including
the 2008 crisis, the COVID-19 shock, and the LIBOR--OIS--SOFR reform.

\medskip
\noindent\textbf{(5) MSGARCH matches implied volatility in tranquil periods and reacts to structural breaks.}
As shown in Section \ref{sec:TotalVar}, MSGARCH and standard GARCH filters track the SRVIX index
and ATM swaption volatility similarly during stable SOFR-era conditions, while
MSGARCH reacts more strongly during genuine regime shifts. This highlights that its
strength lies in identifying structural changes rather than improving short-horizon
forecasting performance.

\medskip
Standard GARCH and GJR-GARCH models remain useful benchmarks for short-run
forecasting and residual diagnostics. However, their instability in long-run variance
limits their interpretability over extended samples. The MSGARCH framework provides
the most internally consistent and economically meaningful representation of long-run
volatility behaviour, especially in markets shaped by evolving collateralisation
practices and benchmark reforms.

}

\section*{Acknowledgements}
The authors thank the anonymous Reviewers for their constructive comments and valuable suggestions, 
which significantly improved the clarity, methodological rigor, and overall quality of this paper.

\section*{Conflict of interest}
In accordance with Taylor \& Francis policy and our ethical obligation as a researchers we state that there is no conflict of interest in this research.

\section*{Declaration of funding}
This work was supported by the Polish Ministry of Science and Higher Education under Grant DWD/6/0551/2022 within 6th edition of the Implementation Doctorate Program.
The Grant funded data purchase used in this article.

\end{document}